\documentclass[aps,pra,twocolumn,superscriptaddress,floatfix,nofootinbib,showpacs,longbibliography]{revtex4-1}
\usepackage[utf8]{inputenc}  
\usepackage[T1]{fontenc}     
\usepackage[british]{babel}  
\usepackage[sc,osf]{mathpazo}
\usepackage[scaled=0.86]{berasans}  
\usepackage[colorlinks=true, citecolor=blue, urlcolor=blue]{hyperref}  
\usepackage{graphicx}
\usepackage{tabularray}
\UseTblrLibrary{amsmath}
\usepackage[babel]{microtype}  
\usepackage{amsmath,amssymb,amsthm,bm,amsfonts,mathrsfs,bbm} 

\usepackage{xspace}  
\usepackage{pgf,tikz}
\usepackage{xcolor}
\usepackage{multirow}
\usepackage{array}
\usepackage{bigstrut}
\usepackage{braket}
\usepackage{color}
\usepackage{natbib}
\usepackage{multirow}
\usepackage{mathtools}
\usepackage{float}
\usepackage{xcolor,colortbl}
\usepackage{physics}
\usepackage{amsmath}
\usepackage{color}
\usepackage{bbold}
\usepackage[justification=justified, format=plain]{subcaption}
\usepackage[justification=raggedright]{caption}

\newcommand{\be}{\begin{equation}}
\newcommand{\ee}{\end{equation}}
\newcommand{\ba}{\begin{eqnarray}}
\newcommand{\ea}{\end{eqnarray}}

\def\>{\rangle}
\def\<{\langle}

\begin{document}
	
\title{Interplay between the Hilbert-space dimension of a control system and the memory induced by a quantum \texttt{SWITCH}}	

\author{Saheli Mukherjee}
\email{mukherjeesaheli95@gmail.com}
\affiliation{S. N. Bose National Centre for Basic Sciences, Block JD, Sector III, Salt Lake, Kolkata 700 106, India}

\author{Bivas Mallick}
\email{bivasqic@gmail.com}
\affiliation{S. N. Bose National Centre for Basic Sciences, Block JD, Sector III, Salt Lake, Kolkata 700 106, India}

\author{Sravani Yanamandra}
\email{sravani.yanamandra@research.iiit.ac.in}
\affiliation{Centre for Quantum Science and Technology}
\affiliation{Center for Security Theory and Algorithmic Research, International Institute of Information Technology,Gachibowli,Hyderabad 500032,India}

\author{Samyadeb Bhattacharya}
\email{samyadeb.b@iiit.ac.in}
\affiliation{Centre for Quantum Science and Technology}
\affiliation{Center for Security Theory and Algorithmic Research, International Institute of Information Technology,Gachibowli,Hyderabad 500032,India}

\author{Ananda G. Maity}
\email{anandamaity289@gmail.com}
\affiliation{Networked Quantum Devices Unit, Okinawa Institute of Science and Technology Graduate University, Onna-son, Okinawa 904-0495, Japan}

\begin{abstract}
Several recent studies have demonstrated the utility of the quantum \texttt{SWITCH} as an important resource for enhancing the performance of various information processing tasks. In a quantum \texttt{SWITCH}, the advantages appear significantly due to the coherent superposition of alternative configurations of the quantum components which are controlled by an additional control system. Here we explore the impact of increasing the Hilbert-space dimension of the control system on the performance of the quantum \texttt{SWITCH}. In particular, we focus on a quantifier of the quantum \texttt{SWITCH} through the emergence of non-Markovianity and explicitly study their behavior when we increase the Hilbert-space dimension of the control system. We observe that increasing the Hilbert-space dimension of the control system leads to the corresponding enhancement of the non-Markovian memory induced by it. Our study demonstrates how the dimension of the control system can be harnessed to improve the quantum \texttt{SWITCH}-based information processing or communication tasks.
\end{abstract}
\maketitle

\section{Introduction}
Shannon's pioneering work established the structure of information theory, underpinning modern communication technology \cite{Shannon48}. In Shannon's theory, devices adhere to the principle of classical physics, yet at a fundamental level, these classical laws are mere approximations of the laws of quantum physics. Over time, there has been a significant focus on investigating communication protocols pertaining to the transfer of quantum data. The transmission of quantum data offers promising advantages in various information processing and communication tasks compared to its classical counterpart. This leads to the emergence of the quantum Shannon theory \cite{Chuang00,Wilde13}. 

However, even in the traditional approach of quantum Shannon theory, there is an implicit assumption that information carriers adhere to a predetermined and fixed causal framework. In contrast, quantum theory accommodates a scenario wherein information carriers are organized in a coherent superposition of alternative orders. In the past few years, much interests are devoted to adopt the phenomenon known as indefinite causal order \cite{Chiribella13,chiribella2012perfect,Oreshkov12} to explore its potential advantages in various information processing tasks. In an indefinite causal order framework, an additional ancillary system is used to control the sequence in which the channels act. For instance, consider two channels denoted by $\mathcal{N}_1$ and $\mathcal{N}_2$. Within this framework, the ancillary qubit determines the sequence in which these channels operate. If the control qubit is initially set to the state $\ket{0}$, it results in one specific order of channel action, namely, $\mathcal{N}_2 \circ \mathcal{N}_1$. Conversely, if the control qubit is prepared in the state $\ket{1}$, the channels will operate in the order $\mathcal{N}_1 \circ \mathcal{N}_2$. However, if the control qubit is prepared in a state $\ket{+}=\frac{\ket{0}+\ket{1}}{\sqrt{2}}$ state, then the two configurations will exist in superposition \cite{Chiribella13}. Such indefinite configuration is shown to be valuable resource for testing the properties of quantum channels \cite{chiribella2012perfect}, increasing communication \cite{Ebler18,Chiribella21,Banik21,akibue2017entanglement} and computation rates \cite{Araujo15}, reducing communication complexities \cite{Guerin18}, improving measurement precision \cite{Zhao20}, exploring thermodynamic advantages \cite{Tamal20,Vedral20,simonov2022activation,liu2022thermodynamics}, and numerous other applications \cite{maity2024activating,Mukhopadhyay19,Ghosal22,van2023device,yoshida2023reversing,quintino2019reversing,bavaresco2021strict}. Several experiments have confirmed these advantages as well \cite{Procopio15,Rubino17,rubino2021experimental,Goswami18(1),goswami2020experiments}.

Despite the numerous potential applications of the concept of indefinite causal order, there remain several aspects of the quantum \texttt{SWITCH} that necessitate thorough investigations. Specifically, a comprehensive investigation into the fundamental resources underpinning the advantages of the quantum \texttt{SWITCH} is yet to be conducted. Recent discoveries highlight the importance of maintaining coherence in the control qubit during both the preparation and measurement steps to uncover the advantages associated with a quantum \texttt{SWITCH} \cite{anand2023emergent}. However, our study primarily aims to investigate another fundamental quantum property that plays a pivotal role in the operation of the quantum \texttt{SWITCH}. Here we show that the fundamental enhancement of the advantages behind the action of quantum \texttt{SWITCH} also hinges on the Hilbert-space dimension of the control system. To investigate this explicitly, we particularly explore how the non-Markovian memory of a channel can be enhanced by increasing the Hilbert-space dimension of the control system. Specifically, we employ the recently introduced quantifier of quantum \texttt{SWITCH}, obtained from the induced non-Markovianity \cite{anand2023emergent}, resulting from the effective dynamics of the quantum \texttt{SWITCH} operation. We examine the behavior of the depolarizing dynamics under the action of quantum \texttt{SWITCH} by increasing the dimension of the control system and compute the quantifier after its action on the dynamics. We adopt the cyclic \texttt{SWITCH} model as proposed by Chiribella {\it et al.} \cite{chiribella2021quantum} and illustrate the fact that enhancing the dimension of the control system will lead to the increment of the quantifier of quantum \texttt{SWITCH}, as evidenced by the induced non-Markovian memory. The formalism of cyclic \texttt{SWITCH} has also been considered earlier in other specific applications of indefinite causal order e.g., to study the classical capacity of a quantum channel \cite{sazim2021classical,procopio2019communication,procopio2020sending}, to analyze the extractable work from a quantum state \cite{simonov2022activation} and many more \cite{nie2022quantum}.

The rest of the paper is organized as follows. In section \ref{s2}, we provide a brief overview of indefinite causal order using the quantum \texttt{SWITCH}, along with an introduction to channels operating in a superposition of cyclic orders. Additionally, we also discuss about the quantifier of the quantum \texttt{SWITCH} through induced non-Markovian memory. In section \ref{s3}, we explicitly study the action of the quantum \texttt{SWITCH} for different dimensions of the control system for depolarizing dynamics. A generalization of the \texttt{SWITCH} action to an arbitrary dimension $n$ of the control system is also presented here. Further, we study the performance of the full quantum-\texttt{SWITCH} consisting of $n!$ dimensions and the cyclic \texttt{SWITCH} model consisting of $n$ dimensions, while keeping the number of channels fixed in each case. Finally, in section \ref{s5}, we summarize our main findings along with some future perspectives.

\section{Preliminaries}\label{s2}
In this section, we shall introduce essential tools that form the basis of our paper. We shall also provide a detailed discussion of the scenario involving indefinite causal order, within which the channels are investigated. We use the terminology ``target state'' for the system of interest and ``control state'' for the ancillary system. Also, we shall consistently employ the standard notations and terminologies commonly utilized in the field of quantum information theory throughout the paper.

\subsection{Quantum \texttt{SWITCH}}
\begin{figure}[ht]
\includegraphics[width=.45\textwidth]{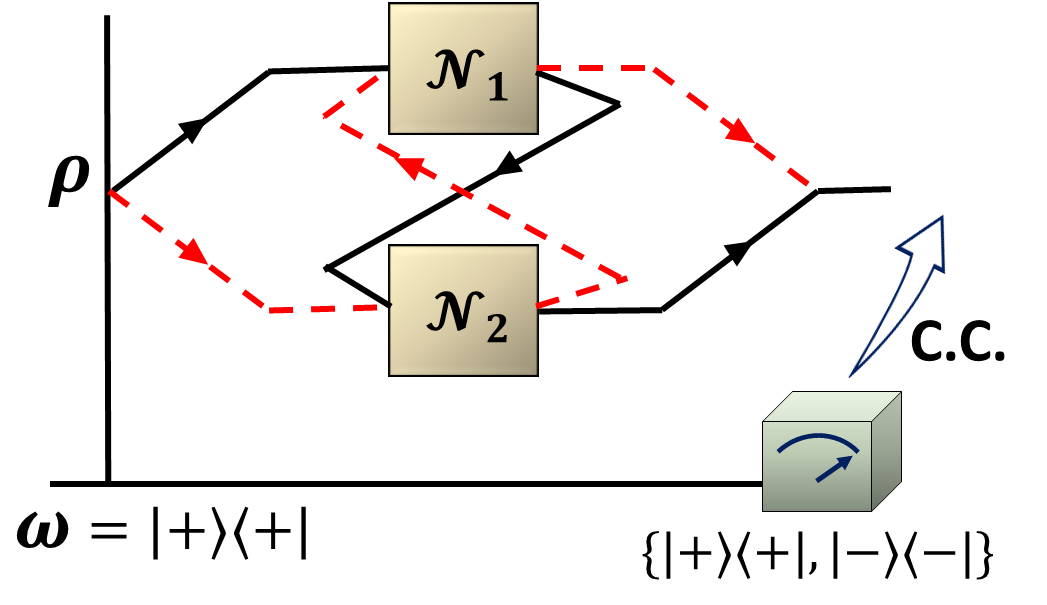}
\caption{The solid black line represents the \texttt{SWITCH} action when the control qubit is initialised in $\omega=\ket{0}\bra{0}$, i.e., the target quantum state $\rho$ first traverses through the channel $\mathcal{N}_{1}$ and then through $\mathcal{N}_{2}$. The dashed red line represents the \texttt{SWITCH} action when the control qubit is initialised in $\omega=\ket{1}\bra{1}$, i.e., the target quantum state $\rho$ first traverses through the channel $\mathcal{N}_{2}$ and then through $\mathcal{N}_{1}$. If $\omega=\ket{+}\bra{+}$, the state $\rho$ traverses through the effective channel made of superposition of $\mathcal{N}_{1} \circ \mathcal{N}_{2}$ and $\mathcal{N}_{2} \circ \mathcal{N}_{1}$. Finally, the control qubit is accessed through a measurement performed in the coherent basis.}\label{Pic1}
\centering
\end{figure}

Quantum channels are the main pathways for sending quantum information to a distant location. A channel $\mathcal{N}$ is nothing but a completely positive and trace preserving (CPTP) map that acts on the linear operator space, i.e., $\mathcal{N} : \mathcal{L}(\mathcal{H}_1) \rightarrow  \mathcal{L}(\mathcal{H}_2)$. The action of the quantum channel can be represented by the Kraus representation $\mathcal{N} (\rho) = \sum_{i}^{} E_{i}\rho{E_{i}^{\dagger}}$ with $\sum_{i} E_{i}^{\dagger} E_{i}= \mathbb{1}$ and $\{E_{i}\} $ are the Kraus operators. Two or more channels can be used either in series or in parallel combination. For two channels $\mathcal{N}_1$ and $\mathcal{N}_2$, the parallel and the series combinations are respectively denoted by the operations $E_{i_{1}} \otimes E_{i_{2}}$ and $E_{i_{1}} \circ E_{i_{2}}$, where $\{E_{i_{1}}\}$ and $\{E_{i_{2}}\}$ are the Kraus operators for the channels $\mathcal{N}_1$ and $\mathcal{N}_2$ respectively. In case of a definite causal order, the channels act in a definite way, i.e., either $\mathcal{N}_{1} \circ \mathcal{N}_{2}$ or  $\mathcal{N}_{2} \circ \mathcal{N}_{1}$ takes place. However, a control qubit can be used to decide the order of the action of the channels leading to the emergence of the phenomenon called indefinite causal order. If the control qubit is in a state $\ket{0}$, then the target system first transmits through the channel $\mathcal{N}_{1}$ and then the channel $\mathcal{N}_{2}$.  This is represented by the Kraus operator, $E_{i_{2}}E_{i_{1}} \otimes \ket{0}\bra{0}$. On the other hand, if the control qubit is in a state $\ket{1}$, then channels act in a reverse way, i.e., $\mathcal{N}_{2}$ is followed by $\mathcal{N}_{1}$ and the corresponding Kraus operator is represented by $E_{i_{1}}E_{i_{2}} \otimes \ket{1}\bra{1}$. Quantum theory also allows coherent superposition of two such configurations with the joint Kraus operator
\begin{equation}\label{2_Kraus}
S_{i_{1}i_{2}}=E_{i_{2}}E_{i_{1}}\otimes \ket{0}\bra{0}+E_{i_{1}}E_{i_{2}}\otimes \ket{1}\bra{1}.
\end{equation}
If the initial state of the target and the control are $\rho$ and $\omega$ respectively, then the evolution of the joint target-control state under the \texttt{SWITCH} action is given by 
\begin{equation}
   W(\mathcal{N}_1, \mathcal{N}_{2})(\rho \otimes \omega) = \sum_{i_{1}i_{2}} S_{i_{1}i_{2}}(\rho \otimes \omega) S_{i_{1}i_{2}}^{\dagger}.
\end{equation}
Now, when the initial state of the control is prepared in  $\omega = \ket{+}\bra{+}$ state and finally the control qubit is measured in $\{\ket{+},\ket{-}\}$ basis, then the target state reduces to
\begin{equation}\label{2_SWITCH}
\Lambda ^{S^{(2)}}(\rho) =\langle \pm | W(\mathcal{N}_1, \mathcal{N}_{2})(\rho \otimes \ket{+}\bra{+})|\pm\rangle 
\end{equation}
where $\Lambda ^{S^{(2)}}(\rho)$ is the map corresponding to the superchannel for a $2$-dimensional control. Here, it may be noted that in quantum \texttt{SWITCH} mechanisms, the advantages are primarily showcased in conditional or post-selected states determined by the measurement outcomes whereas tracing out the control destroy the indefinite causal order yielding no advantage. In principle, there can be different choices for the initial state of control and for the post-selection corresponding to different measurement outcomes. However, to fully utilize the power of quantum \texttt{SWITCH}, we need to optimize over all such possible choices. The action of quantum \texttt{SWITCH} as a coherent superposition of channel order is depicted in Fig.~\ref{Pic1}.

\subsection{Quantum channels in a superposition of cyclic orders }
As discussed in the previous subsection, quantum \texttt{SWITCH} can be used to construct superposition of two causal orders where two channels are taken as inputs. However, driven by the motivation to investigate the impact of higher-dimensional control in quantum \texttt{SWITCH}, as opposed to utilizing just two channels, we aim to employ $n$ channels, where $n$ can be arbitrarily large. Within this extended framework accommodating $n$ channels, there are a total of $n!$ distinct permutations of channel orderings, and these different superpositions of channel arrangements can be achieved by employing a control system with $n!$ dimensions. The joint Kraus operator will be
\begin{equation}
S_{i_{1}i_{2}....i_{n}}=\sum_{m=0}^{l-1}P_{m}(K_{i_1},K_{i_2},...,K_{i_n}) \otimes \ket{m}\bra{m}
\end{equation}
where $l\in[2,n!]$ and $P_{m}(K_{i_1},K_{i_2},...,K_{i_n})$ denotes the different permutations of the channels having Kraus operators $\{K_{i_1}\},\{K_{i_2}\},....,\{K_{i_n}\}$ for $n$ channels respectively.

\begin{figure}[ht]
\includegraphics[width=.5\textwidth]{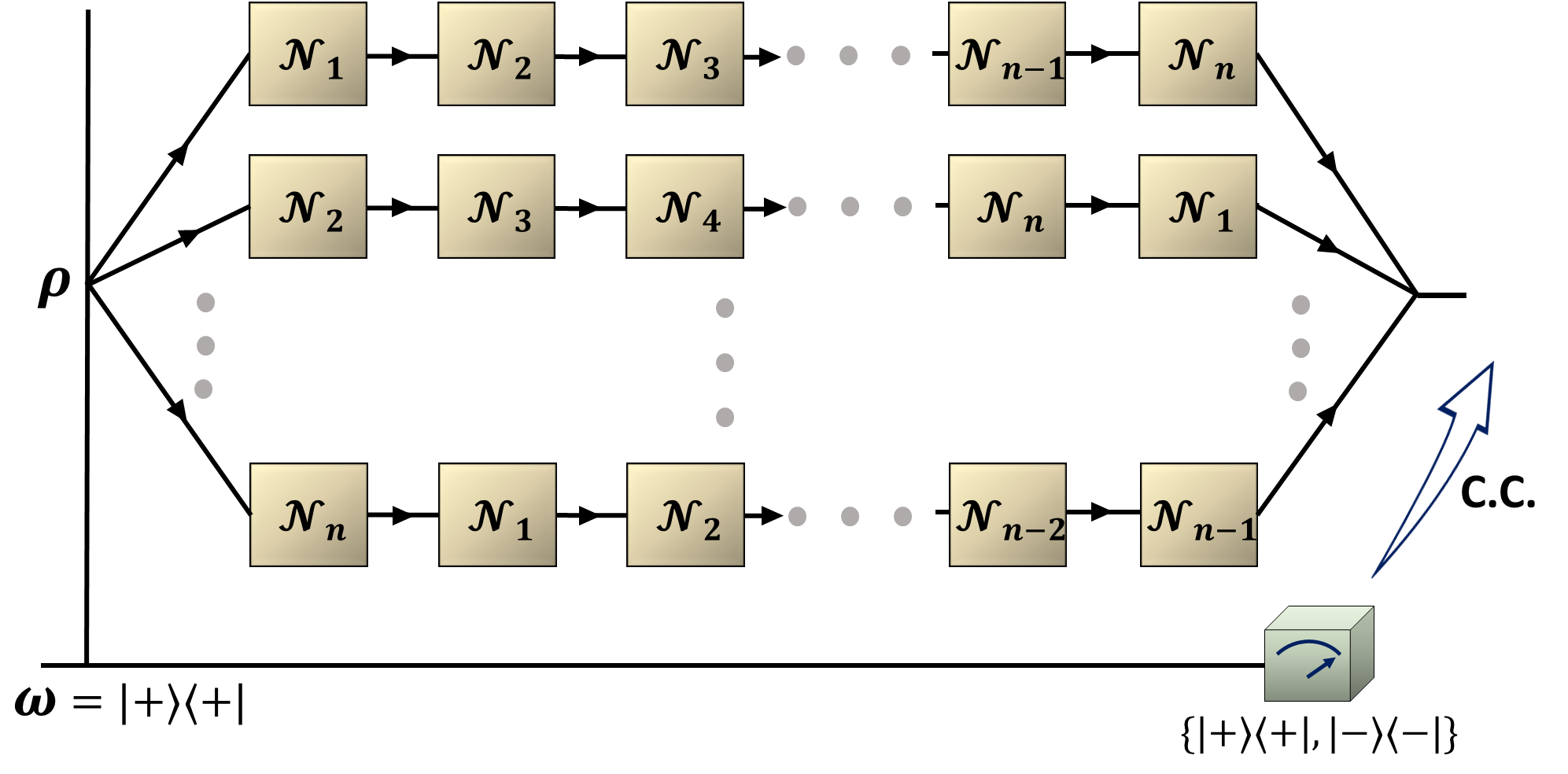}
\caption{When the control is initialised in $\omega=\ket{0}\bra{0}$, the target state $\rho$ first traverses though the channel $\mathcal{N}_{1}$, then subsequently through the channels $\mathcal{N}_{2}$, $\mathcal{N}_{3}$,...,$\mathcal{N}_{n}$, i.e.,the channels acts as $\mathcal{N}_{n} \circ \mathcal{N}_{n-1} \circ .....\circ \mathcal{N}_{2} \circ \mathcal{N}_{1}$. Similarly, when the control is initialised in $\omega=\ket{1}\bra{1}$, the target state $\rho$ first traverses though the channel $\mathcal{N}_{2}$, then subsequently through the channels $\mathcal{N}_{3}$, $\mathcal{N}_{4}$,...,$\mathcal{N}_{n}$, $\mathcal{N}_{1}$, i.e.,the channels acts as $\mathcal{N}_{1} \circ \mathcal{N}_{n} \circ .....\circ \mathcal{N}_{3} \circ \mathcal{N}_{2}$. For $\omega=\ket{n-1}\bra{n-1}$, the action of the channels is as $\mathcal{N}_{n-1} \circ \mathcal{N}_{n-2} \circ .....\circ \mathcal{N}_{1} \circ \mathcal{N}_{n}$. The channels thus act in a cyclic order. However, preparing the control system in coherent basis will lead to the superposition of all such cyclic orders. Finally, the control is accessed through a measurement performed in the generalized coherent basis.}\label{Pic2}
\centering
\end{figure}

On the other hand for the case of cyclic permutations, the joint Kraus operator can be written as, 
 \begin{equation}
\begin{split}
S_{i_{1}i_{2}i_{3}....i_{n}}=& 
 K_{i_{n}} K_{i_{n-1}}....K_{i_{2}} K_{i_{1}} \otimes \ket{0}\bra{0}\\
&+K_{i_{1}}K_{i_{n}}....K_{i_{3}}K_{i_{2}} \otimes \ket{1}\bra{1} \\ &+K_{i_{2}}K_{i_{1}}K_{i_{n}}....K_{i_{3}} \otimes \ket{2}\bra{2}+....\\ &+K_{i_{n-1}}K_{i_{n-2}}....K_{i_{1}}K_{i_{n}} \otimes \ket{n-1}\bra{n-1}. \label{superchannel}
\end{split}
\end{equation}
After the \texttt{SWITCH} action when the control system is measured in $\{\ket{+},\ket{-}\}$ basis (where generalized $\ket{+}=\frac{\ket{0}+\ket{1}+.....+\ket{n-1}}{\sqrt{n}}$ and $\ket{-}$ denotes the state(s) orthogonal to $\ket{+}$ ), the updated target state corresponding to the $'+'$ outcome will become,
\begin{equation}
\begin{split}
     \rho^{(n)}(t)=& \Lambda ^{S^{(n)}}(\rho(0)) \\
     =& \langle +|S_{i_{1}i_{2}....i_{n}}(\rho(0)\otimes \ket{+}\bra{+})S_{i_{1}i_{2}....i_{n}}^{\dagger}|+\rangle. \label{switch3}
  \end{split}
\end{equation}
From now onwards, in the cyclic \texttt{SWITCH} scenario, we shall denote the action of $n$-th dimensional control system as $n$-\texttt{SWITCH} and hence by our definition the conventional quantum \texttt{SWITCH} will be denoted as $2$-\texttt{SWITCH}. Here it is important to mention that in Ref. \citep{sazim2021classical}, the authors also concentrated on cyclic orders, noting that the imminent mixing of cyclic and non-cyclic orders may led to a decrease in Holevo information, and hence cyclic orders are always more advantageous than non-cyclic ones for a fixed number of channels and control dimension. For this reason, we will also adopt the cyclic \texttt{SWITCH} formalism mostly for our further analysis in the subsequent sections. The overall action of the $n$-\texttt{SWITCH} as a coherent superposition of $n$ channels in cyclic order is depicted in Fig.~\ref{Pic2}.
Eq.\eqref{switch3} can also be written in an alternative form
\begin{equation}
     \frac{\sum_{i_{1}i_{2}....i_{n}} M_{i_{1}i_{2}....i_{n}} \rho(0) M_{i_{1}i_{2}....i_{n}}^{\dagger}}{\Tr(\sum_{i_{1}i_{2}....i_{n}} M_{i_{1}i_{2}....i_{n}} \rho(0) M_{i_{1}i_{2}....i_{n}}^{\dagger})} \label{switchkraus}
  \end{equation}
  with \\$M_{i_{1}i_{2}....i_{n}} = \frac{K_{i_{n}}....K_{i_{2}}K_{i_{1}}+ K_{i_{1}}K_{i_{n}}....K_{i_{2}}+....+K_{i_{n-1}}K_{i_{n-2}}....K_{i_{1}}K_{i_{n}}}{n}$.
  The detailed proof for this equivalence for the case of $2$-\texttt{SWITCH} is given in appendix \ref{A}. In the rest of the paper, we shall use this formula (Eq.\eqref{switchkraus}) to calculate the updated target state after the \texttt{SWITCH} action for any values of $n$. It may be noted here that we can also consider the post-selection corresponding to the other outcome and analogously define the evolution (Kraus) operator using the same approach.
\subsection{Quantifier of quantum \texttt{SWITCH} through induced non-Markovianity}
Inspired by the extensive utility of the quantum \texttt{SWITCH} in various quantum information processing, communication, and computational applications, very recently dynamical quantification of quantum \texttt{SWITCH} has been proposed \citep{anand2023emergent}. In particular, the authors have explored a specific type of quantum memory that arises when examining the dynamics associated with the operation of the quantum \texttt{SWITCH}. More specifically, a connection has been established between the memory induced by the quantum \texttt{SWITCH} and the information loss that occurs due to the interaction between the target system and the environment through the switched channel. The quantification of this information loss is determined by the disparity in the distance between the two states prior to and subsequent to the switching operation. Their findings demonstrate that, for the time-averaged states, an increase in quantum \texttt{SWITCH}-induced memory corresponds to a decrease in information loss. This indicates that the distance between the two states evolving under the action of quantum \texttt{SWITCH} increases which can be attributed to the backflow of information from the control to the system and hence, corresponds to a non-Markovian dynamics \cite{RHPreview,BLPreview,Vegareview,laine2010measure,rivas2010entanglement,Maity20,mallick2024assessing}. In open quantum systems, the system interacts with the environment and depending on the nature of the evolution (memoryless or not), the dynamics can be one of the two types-- Markovian or non-Markovian. A dynamics is called Markovian if it is completely positive and satisfies the divisibility rule
\begin{equation}
\Lambda (t,t_0)= \Lambda (t,s) \circ \Lambda (s,t_0) \label{divisibility}
\end{equation}
for $t \ge s \ge t_{0}$. In Markovian dynamics, the environment does not retain any memory of past interactions with the system contrary to a non-Markovian dynamics, where the memory effect is evident. Utilizing the example of a Markovian qubit completely depolarizing channel, in Ref. \cite{anand2023emergent} it has been shown that while the original channel is Markovian, the switched channel exhibits certain non-Markovian effects stemming from the memory induced by the \texttt{SWITCH} action which is quantified by some non-Markovian quantifier. Specifically, the authors consider the RHP measure (named after Rivas, Huelga, and Plenio) \citep{rivas2010entanglement} and the BLP measure (proposed by Breuer, Laine, and Piilo) \cite{breuer2009measure} of non-Markovianity. Note that since the \texttt{SWITCH}-induced memory turns out to be non-Markovian, we can consider any measure of non-Markovianity. However, in this work, we shall mainly focus on the RHP measure. RHP measure is based on the complete positive (CP) divisibility of a dynamical map and can formally be defined as 
\begin{equation}
 \mathcal{I}= \int_{0}^{\infty} g(t) dt
\end{equation} where
\begin{equation}
  g(t) = \lim_{ \Delta t\to 0} \frac{||(\mathbb{1} \otimes \Lambda _{\Delta t}) \ket{\phi}\bra{\phi}||_{1}-1}{\Delta t},
\end{equation}
$\Lambda_{\Delta t}= (\mathbb{1}+ \Delta t\mathcal{L}_{t})$ for a Lindblad type dynamics and $\ket{\phi}\bra{\phi}$ is a maximally entangled state. The normalized measure is thus given by $\frac{\mathcal{I}}{\mathcal{I}+1}$.
By definition, for a Markovian dynamics, if the map is CP divisible and if it satisfies the trace-preserving condition, then $g(t)=0$. However, if at any point of time, the CP divisibility breaks, then $g(t)>0$ and the dynamics is non-Markovian. Formally, $\int_{0}^\infty g(t) dt$ is a measure of non-Markovianity.

In this study, we start with Markovian channels (or dynamical maps) and investigate the action of quantum \texttt{SWITCH} on those channels. Starting from an initial state, the final target state is obtained after the action of the quantum \texttt{SWITCH} on these channels, followed by a measurement on the control system, and then post-selection corresponding to the particular outcome. Alternatively, one can investigate the particular dynamics leading to the same final state from the same initial state. We analyze this dynamics and demonstrate that they exhibit non-Markovian traits although the initial dynamical maps (in absence of the quantum \texttt{SWITCH} action) are considered to be Markovian. This suggests that the non-Markovian memory is induced due to the \texttt{SWITCH} operation. In the next section, we shall show that this manifestation of non-Markovian behavior resulting from the \texttt{SWITCH}-induced memory becomes more pronounced as the Hilbert-space dimension of the control system increases.

\section{Action of the quantum \texttt{SWITCH} on depolarizing dynamics} \label{s3}
To investigate the action of \texttt{SWITCH} with a control system of arbitrary dimension $(n \ge 2)$,  we begin by examining the basic qubit depolarizing channel. A qubit system interacts with the environment, and the system's evolution is determined by an invertible map characterized by a Lindblad-type generator $\mathcal{L}_{t}$ i.e.,
\begin{equation}
    \frac{d\rho}{dt}={\mathcal{L}_{t}}(\rho(t))
\end{equation}
where ${\mathcal{L}_{t}}(\rho(t))=\sum_{i}\Gamma_{i}(t)(A_{i}\rho A_{i}^{\dagger}-\frac{1}{2}\{A_{i}^{\dagger}A_{i},\rho\})$
with $\Gamma_{i}(t)$ and $A_{i}$ being the Lindblad coefficients and the Lindblad operators respectively. The action of the map is
\begin{equation}
    \Lambda(\rho)=\exp(\int_{0}^{t} \mathcal{L}_{s}\,ds)
\end{equation}
The master equation is then given by
\begin{equation}
\frac{d\rho(t)}{dt}=\sum_{i=1}^{3}\gamma_{i}(t)[\sigma_{i}\rho(t)\sigma_{i}-\rho(t)] \label{master}
\end{equation}
where $\gamma_{i}(t)$ are the Lindblad coefficients and $\sigma_{i} (i=1,2,3)$ are the Pauli matrices. The state after the channel action is represented by
\begin{equation}
\Lambda(\rho)=\rho(t)= \begin{bmatrix}
\rho_{11}(t) & \rho_{12}(t) \\
\rho_{21}(t) & \rho_{22}(t) \label{state} \\
\end{bmatrix}
\end{equation}
where,
\begin{equation}
\begin{split}
      &\rho_{11}(t) = \rho_{11}(0)\left(\frac{1+e^{-2\zeta_{1} (t)}}{2}\right) +  \rho_{22}(0) \left(\frac{1-e^{-2 \zeta_{1} (t)}}{2}\right),\\
      & \rho_{22}(t) = 1- \rho_{11}(t) , \hspace{0.5cm}  \rho_{12/21}(t) = \rho_{12/21}(0) e^{-2\zeta_{2} (t)}, \nonumber
   \end{split}
\end{equation}
with 
\begin{equation}
    \begin{split}
    &    \zeta_{1}( t) =  \int_{0}^{t} [\gamma_1 (s) + \gamma _2 (s)] \,ds \\
     &    \zeta_{2}( t) =  \int_{0}^{t} [\gamma_2 (s) + \gamma _3 (s)] \,ds. \nonumber
     \end{split}
\end{equation}
For a completely depolarizing channel, the Kraus operators are given by 
\begin{equation}
\begin{split}
& K_1= \sqrt{{A_2}(t)}\begin{bmatrix}
0 & 1 \\
0 & 0 \\
\end{bmatrix}, ~~K_2= \sqrt{{A_2}(t)}\begin{bmatrix}
0 & 0 \\
1 & 0 \\
\end{bmatrix}, \\
& K_3= \sqrt{\frac{{A_1}(t) +{A_3}(t)}{2}}\begin{bmatrix}
e^{\iota \theta(t)} & 0 \\
0 & 1 \\
\end{bmatrix}, \\
& K_4= \sqrt{\frac{{A_1}(t) -{A_3}(t)}{2}}\begin{bmatrix}
-e^{\iota \theta(t)} & 0 \\
0 & 1 \\
\end{bmatrix}. \label{kraus}
\end{split}
    \end{equation} 
Here,
    \begin{equation}
        \begin{split}
         &   A_1= \frac{1+e^{-2\zeta_{1} (t)}}{2}, A_2= \frac{1-e^{-2\zeta_{1} (t)}}{2}\\
         & A_3= e^{-2\zeta_{2} (t)},  \hspace{0.5cm} \theta(t) = \tan^{-1}\left(\frac{\text{Im} (A_3(t))}{\text{Re} (A_3(t))}\right) =0. \nonumber
        \end{split}
    \end{equation}

The dynamics is Markovian iff $\gamma_{i}(t) \ge 0, \forall i,t$. For simplicity, here, we take $\gamma_1(t) = \gamma_2(t) = \gamma_3(t) = \gamma > 0$. 

Now, let's examine the operation of the quantum \texttt{SWITCH} on this depolarizing dynamics. When the dimension of the control system is $n$, we examine a scenario involving $n$ number of qubit depolarizing channels arranged in a cyclic permutation, each channel being described by the Kraus operators as given in Eq.\eqref{kraus}. If the initial target qubit state of the system is $\rho(0)$, then the state of the target system after the action of quantum \texttt{SWITCH} is $\Lambda^{S^{(n)}}(\rho(0))=\rho^{(n)}(t)=$
\begin{equation}\small
  \setlength{\arraycolsep}{0.5pt}
  \renewcommand{\arraystretch}{0.3}
  \begin{pmatrix}
  A^{(n)}(t) \rho_{11}(0) + B^{(n)}(t) \rho_{22}(0)&C^{(n)}(t) \rho_{12}(0) \\
C^{(n)}(t) \rho_{21}(0) & B^{(n)}(t) \rho_{11}(0) + A^{(n)}(t) \rho_{22}(0)\\  \label{updatedswitchstate}
  \end{pmatrix}
\end{equation}
where we have considered a post-selection of the control system corresponding to the $'+'$ outcome with $\ket{+}=\frac{(\ket{0}+\ket{1}+...+\ket{n-1}}{\sqrt{n}}$. Further, $A^{(n)}(t) = \frac{1+C^{(n)}(t)}{2},  B^{(n)}(t) = \frac{1-C^{(n)}(t)}{2}$ and $C^{(n)}(t)$ are $n$-th order polynomial in $\eta$ (with $\eta=e^{-4\gamma t}$) as derived in appendix \ref{B}. The proof for the linearity of this map is given in appendix \ref{C}. The Master equation for the \texttt{SWITCH} action will be of the form
  \begin{equation}
      \frac{d\rho}{dt}= \sum_{i} \Gamma_{i}^{(n)}(t) [\sigma_{i}\rho(t)\sigma_{i}-\rho] \label{switchmaster}
  \end{equation}
where $\Gamma_{1}^{(n)}(t)=\Gamma_{2}^{(n)}(t)=\Gamma_{3}^{(n)}(t)=\Gamma^{(n)}(t)=-\frac{1}{4}\frac{d}{dt}(\ln C^{(n)}(t))$ \cite{bhattacharya2017exact}. The dynamics after the \texttt{SWITCH} action is said to exhibit non-Markovian behavior iff $\Gamma^{(n)}(t) < 0$ for some $t$. The explicit expression for $C^{(n)}(t)$ and $\Gamma ^{(n)} (t)$ corresponding to the individual dimensions and their comprehensive analysis are formally expressed in the subsequent subsections. For the particular dynamics considered by us, $\Gamma ^{(n)} (t)$ becomes negative after certain time $T_{-}^{(n)}$ (i.e., $\Gamma^{(n)}(t) < 0$ for $t>T_{-}^{(n)}$), and will continue to be negative till $t=\infty$. We designate the moment $T_{-}^{(n)}$ as the characteristic time, denoting the point at which non-Markovianity triggers, i.e., when $\Gamma^{(n)}(T_{-}^{(n)}) = 0$. For the above depolarizing dynamics, the measure of induced non-Markovianity  \cite{anand2023emergent} turns out to be 
\begin{equation}
    g(t)=
     \begin{cases}0 , ~~~~~~~~~~~~~~~~\text{when} \hspace{0.2cm}\gamma(t) \ge 0\\
   -6\Gamma^{(n)}(t), ~~~ \text{when} \hspace{0.2cm} \gamma(t)<0.
    \end{cases}
\end{equation}
Therefore, in order to capture the memory induced by the quantum \texttt{SWITCH}, it is sufficient to integrate over the region where the Lindblad coefficient is negative and hence the \texttt{SWITCH}-induced memory($\mathcal{M}^{(n)}$) reduces to $\int_{t=T_{-}^{(n)}}^\infty -6\Gamma ^{(n)} (t) dt$. The normalized measure is given by $\frac{\mathcal{M}^{(n)}}{1+\mathcal{M}^{(n)}}$. For a more detailed derivation of the \texttt{SWITCH}-induced memory, interested readers are referred to the Ref. \cite{anand2023emergent}.  Note that for the cyclic \texttt{SWITCH} scenario (all the cases considered below), the measure of \texttt{SWITCH}-induced non-Markovianity turns out to be maximum when the initial state of control is prepared at $\ket{+}$ and the control state after the \texttt{SWITCH} action is measured in the $\{\ket{+}, \ket{-}\}$ basis followed by a post-selection corresponding to the $'+'$ outcome. So unless specified otherwise, we shall take this choice of the initial control state and post-selection throughout the paper. The dynamics corresponding to the post-selection resulting in other measurement outcomes can also be derived in a similar fashion. Nevertheless, the Lindblad coefficients ($\gamma_{1}, \gamma_{2}$ and $\gamma_{3}$) can also be considered different in general and this has been studied explicitly in appendix \ref{D}.

\subsection{Action of the $2$-\texttt{SWITCH}}
For the action of $2$-\texttt{SWITCH} (which is the conventional quantum \texttt{SWITCH}), the Kraus operator, $S_{i_{1}i_{2}}$ for the superchannel can be considered as given in Eq.\eqref{2_Kraus}. After the action of the $2$-\texttt{SWITCH}, when the control qubit is measured in the $\{\ket{+},\ket{-}\}$ basis, the effective target state (corresponding to the `$+$' outcome) becomes (from Eq.\eqref{2_SWITCH}) $\rho^{(2)}(t) = \Lambda ^{S^{(2)}}(\rho(0))=\langle +|S_{i_{1}i_{2}}(\rho(0)\otimes \ket{+}\bra{+})S_{i_{1}i_{2}}^{\dagger}|+\rangle$. Now from Eq.\eqref{switchkraus}, $\rho^{(2)}(t)$ can also be written in an alternative form as $\frac{\sum_{i_{1}i_{2}} M_{i_{1}i_{2}} \rho(0) M_{i_{1}i_{2}}^{\dagger}}{\Tr(\sum_{i_{1}i_{2}} M_{i_{1}i_{2}} \rho(0) M_{i_{1}i_{2}}^{\dagger})}$ 
with $M_{i_{1}i_{2}} = \frac{K_{i_{1}}  K_{i_{2}} +  K_{i_{2}}  K_{i_{1}}}{2}$ and $ i_{1},i_{2} \in \{1,2,3,4\}$.
The updated target state is then given by Eq.\eqref{updatedswitchstate} with
\begin{equation}
     C^{(2)}(\eta) = \frac{-(9 \eta^{2}- 2 \eta +1 )}{(3 {\eta^2}- 6\eta -5)}. \nonumber
\end{equation}
The above expression leads to
\begin{equation}
    \Gamma^{(2)}(\eta)=\frac{16 \gamma \eta(3 \eta^{2} +6 \eta -1)}{(9 \eta^{2}-2 \eta +1)(-3 \eta^{2}+6\eta+5)}. \label{gamma2}
\end{equation}
We plot $6\Gamma^{(2)}(t)$ in Fig.~\ref{fig3}. Throughout the paper, $C^{(n)}(t)$ and $\Gamma^{(n)}(t)$ are calculated from $C^{(n)}(\eta)$ and $\Gamma^{(n)}(\eta)$ respectively by using $\eta = e^{-4 \gamma t}$. Now, in order to explicitly calculate the memory induced by the $2$-\texttt{SWITCH}, we first evaluate the characteristics time, $T_{-}^{(2)}$ corresponding to the dynamics of $2$-\texttt{SWITCH}. Since at $t=T_{-}^{(2)}$, $\Gamma^{(2)}(t) = 0$, from Eq.\eqref{gamma2} we get 
$T_{-}^{(2)} = -\frac{1}{4\gamma} \ln{(0.155)}$. This implies $
\mathcal{M}^{(2)}=\frac{3}{2} [\ln C^{(2)}({\infty})-\ln C^{(2)}(T_{-}^{(2)})] =0.385$. The normalized measure is $\frac{0.385}{0.385+1}=0.278$.
 
To comprehend the role of the Hilbert-space dimension of the control system behind the action of quantum \texttt{SWITCH}, in the subsequent subsections, we investigate the effective dynamics when quantum \texttt{SWITCH} is applied with a control system of higher dimension.

\subsection{Action of the $3$-\texttt{SWITCH}}\label{s3b}
For the $3$-\texttt{SWITCH}, the overall Kraus operator can be described as
\begin{equation}
\begin{split}
    S_{i_{1}i_{2}i_{3}}=& K_{i_{3}}K_{i_{2}}K_{i_{1}} \otimes \ket{0}\bra{0}+K_{i_{1}}K_{i_{3}}K_{i_{2}} \otimes \ket{1}\bra{1} \\ &+K_{i_{2}}K_{i_{1}}K_{i_{3}} \otimes \ket{2}\bra{2}.
    \end{split}
\end{equation}
After the action of the $3$-\texttt{SWITCH}, when the control qutrit is measured in the coherent basis, the effective target state corresponding to the `$+$' (here $\ket{+}=\frac{\ket{0}+\ket{1}+\ket{2}}{\sqrt{3}}$) outcome becomes 
\begin{equation}
   \rho^{(3)}(t)= \Lambda ^{S^{(3)}}(\rho(0))=\langle +|S_{i_{1}i_{2}i_{3}}(\rho(0)\otimes \ket{+}\bra{+})S_{i_{1}i_{2}i_{3}}^{\dagger}|+\rangle \nonumber
  \end{equation}
where $i_{1}, i_{2}, i_{3}\in\{1,2,3,4\}$. Above expression for $\rho(t)$ can also be written (using Eq.\eqref{switchkraus}) as
$$\rho^{(3)}(t)=
 \frac{\sum_{i_{1}i_{2}i_{3}} M_{i_{1}i_{2}i_{3}} \rho(0) M_{i_{1}i_{2}i_{3}}^{\dagger}}{\Tr(\sum_{i_{1}i_{2}i_{3}} M_{i_{1}i_{2}i_{3}} \rho(0) M_{i_{1}i_{2}i_{3}}^{\dagger})}$$
where $M_{i_{1}i_{2}i_{3}} = \frac{K_{i_{3}}K_{i_{2}}K_{i_{1}} + K_{i_{1}}K_{i_{3}}K_{i_{2}}+K_{i_{2}}K_{i_{1}}K_{i_{3}}}{3}$. After simplification and comparing with Eq.\eqref{updatedswitchstate} we get 
\begin{equation}
   C^{(3)}(\eta) = \frac{7\eta^{3}-\eta^{2}- \eta +1 }{3( {-\eta^3}+ \eta^{2}+\eta +1 )}. \nonumber
\end{equation}
This implies
\begin{equation}
  \Gamma^{(3)}(\eta)=\frac{-2 \gamma \eta(-3\eta^{4}-6\eta^{3}-12\eta^{2} +2 \eta +1)}{(-7\eta^{6}+8\eta^{5}+7\eta^{4}+4\eta^{3}-\eta^{2}+1)}. \label{gamma3}
\end{equation}
In order to explore the dynamical behavior reflecting the impact of the quantum \texttt{SWITCH}, we plot $6 \Gamma^{(3)}(t)$ with $t$ in Fig.~\ref{fig3}. 

With the aim of explicitly calculating the \texttt{SWITCH}-induced memory, we first compute the characteristics time $T_{-}^{(3)}$ for the $3$-\texttt{SWITCH}.
Since at $t=T_{-}^{(3)}$, $\Gamma^{(3)}(t) = 0$, then from Eq.\eqref{gamma3}, we get $T_{-}^{(3)} = -\frac{1}{4\gamma} \ln{(0.342)}$ implying $\mathcal{M}^{(3)}=\frac{3}{2} [\ln C^{(3)}(\infty)-\ln C^{(3)}(T_{-}^{(3)})]=0.821$. Hence, the normalized measure comes out to be $\frac{0.821}{0.821+1}=0.451$.
Figure \ref{fig3} clearly illustrates that after the action of the $3$-\texttt{SWITCH}, the non-Markovian behavior emerges and seems to be even more pronounced compared to the $2$-\texttt{SWITCH}, despite both starting from the same initial Markovian channel.

\subsection{Action of the $n$-\texttt{SWITCH}} 
With the aim of analyzing the action of the $n$-\texttt{SWITCH}, the Kraus operator for the superchannel can be expressed in the form of Eq.\eqref{superchannel}.
Following the analysis given in section \ref{s2}, after the action of the $n$-\texttt{SWITCH} (when the control system is measured in the superposition basis), the effective target state (corresponding to the `$+$' outcome) becomes
\begin{equation}
\begin{split}
   \rho^{(n)}(t) &=  \Lambda ^{S^{(n)}}(\rho(0))\\ &=\langle +|S_{i_{1}i_{2}....i_{n}}(\rho(0)\otimes \ket{+}\bra{+})S_{i_{1}i_{2}....i_{n}}^{\dagger}|+\rangle \nonumber
   \end{split}
  \end{equation}
where $i_{1},i_{2},...,i_{n}\in\{1,2,3,4\}$.
Above $\rho^{(n)}(t)$ can also be expressed in terms of Eq.\eqref{switchkraus}. Upon simplification and comparing with Eq.\eqref{updatedswitchstate} we get
\begin{equation}
    C^{(n)}(\eta)=\frac{-[(5n-1)\eta^{n}-2 \sum_{i=1}^{n-1}\eta^{i} +(n-1)]}{3(n-1)\eta^{n}-6\sum_{j=1}^{n-1}\eta^{j}-(n+3)} \nonumber
\end{equation}
and
\begin{equation}
  \Gamma^{(n)}(\eta) = 
         \frac{a_{1}b_{2}-b_{1}a_{2}}{a_{2}b_{2}}
\end{equation}
where $a_{1}=-24\gamma\sum_{k=1}^{n-1}k \eta^{k}+12n(n-1)\gamma \eta^{n}$,\\
  $a_{2}= 4 [(n+3)+6\sum_{k=1}^{n-1} \eta^{k} -3(n-1) \eta^{n}]$ ,\\
  $b_{1}=8 \gamma \sum_{k=1}^{n-1}k \eta^{k} - 4 n (5n-1) \gamma \eta^{n}$, \\ 
  $b_{2}=4[(n-1)-2\sum_{k=1}^{n-1}\eta^{k}+(5n-1)\eta^{n}]$. \\

  \begin{figure}[ht]
\includegraphics[width=.45\textwidth]{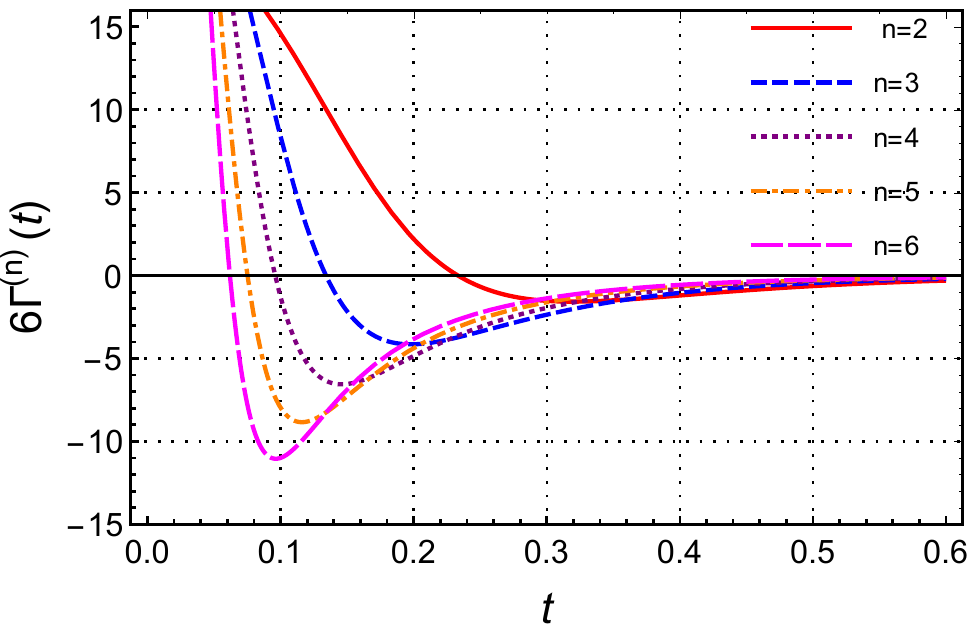} 
\caption{Representation of the \texttt{SWITCH}-induced measure of non-Markovianity in RHP sense for $n$-\texttt{SWITCH} in the region where $\Gamma^{(n)}(t) <0$ and $\gamma=2$. The region where $\Gamma^{(n)}(t) <0$ indicates the emergent non-Markovianity due to the \texttt{SWITCH} action.}\label{fig3}
\centering
\end{figure}

\noindent\textbf{Analysis of the $n$-\texttt{SWITCH} action:}
Fig.~\ref{fig3} displays the combined plots illustrating the \texttt{SWITCH}-induced non-Markovianity of cyclic $n$-\texttt{SWITCH} for different $n$. From Fig.~\ref{fig3}, it is evident that after the action of quantum \texttt{SWITCH}, the effective dynamics show non-Markovian traits. The figure also clearly indicates that after a certain characteristic time the information backflow triggers. Furthermore, as $n$ increases, this effect becomes more prominent, leading to the manifestation of enhanced non-Markovian memory induced by the quantum \texttt{SWITCH}.

To validate this further, we proceed to explicitly compute the memory induced by the quantum $n$-\texttt{SWITCH} with $n$. The quantification of the \texttt{SWITCH}-induced memory can be defined as $\mathcal{M}^{(n)}=\int_{T_{-}^{(n)}}^{\infty} -6\Gamma^{(n)}(t) \,dt$. Upon simplification this expression reduces to $\frac{3}{2}[\ln{C^{(n)}}(\infty)-\ln{C^{(n)}}(T_{-}^{(n)})]$. This indicates that explicit computation of the quantifier requires evaluating the value of the function $\ln C^{(n)}(t)$ at both $t=\infty$ and $t=T_{-}^{(n)}$. However, determining $T_{-}^{(n)}$ involves solving the equation, $\Gamma^{(n)}(t)=0$, where the roots of this $n$-th order polynomial provide the value of $T_{-}^{(n)}$. Hence, obtaining an analytical expression for arbitrary $n$ becomes cumbersome. The values of  the characteristic time and the normalized measure for the \texttt{SWITCH}-induced memory upto $n=15$ is given in Table \ref{table1} in appendix \ref{E}. 

\begin{figure}[ht]
\includegraphics[width=.45\textwidth]{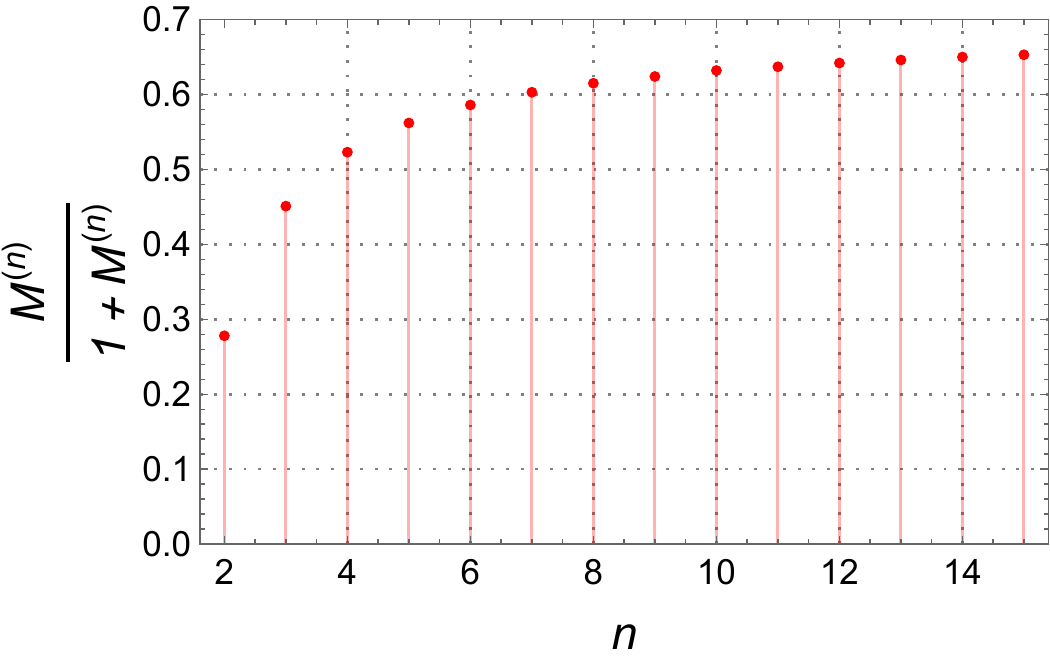} 
\caption{Plot showing the dependence of the quantifier of the quantum \texttt{SWITCH} on $n$.}\label{fig4}
\centering
\end{figure}
Fig.~\ref{fig4} shows the plot of the normalized measure of the \texttt{SWITCH}-induced memory for different values of $n$. It is clear from the graph that the quantifier of the quantum \texttt{SWITCH} increases with $n$ indicating that the augmentation of the control system's dimension enhances the performance of the quantum \texttt{SWITCH}. However, one may still ask whether the increase in the memory induced by quantum \texttt{SWITCH} can solely be attributed to the increase in the dimension of control since the number of channels also increases along with the increase in the dimension of the control. To discard this possibility, here we also evaluate the memory induced by quantum \texttt{SWITCH} by increasing the Hilbert-space dimension $(n)$ of the control system while keeping the number of channels $(X)$ fixed.

For all the cases considered so far, the dimension of the control has been taken to be same as the number of channels. We present an explicit case study of $n \le X$ for several $n$ and $X$ in appendix \ref{F}. The values of the normalized measure for the \texttt{SWITCH}-induced memory for $n \le X$ are summarized in Table \ref{table2} (for explicit calculation we refer to appendix \ref{F}). Our study reveals that the \texttt{SWITCH}-induced memory always increases as $n$ increases for a fixed $X$, as evident from Table \ref{table2}. This demonstrates that increasing the dimension of the control plays a crucial role for the enhancement of the \texttt{SWITCH} performance.

Another important observation that can be inferred from our study is that the performance of the quantum \texttt{SWITCH} may also depend on the number of channels used. This is also confirmed by the data presented in Table \ref{table2}.
\begin{table}[h!]
\begin{tabular}{| c| c| c| c |c |c |}
\hline
  $n$ &  $X$ & $T^{(n),X}_{-}$  & Measure ($\mathcal{M}_{X}^{(n)})$ &  Normalized measure $(\frac{\mathcal{M}_{X}^{(n)}}{1+\mathcal{M}_{X}^{(n)}})$\\
 
 \hline
 $2$  & $3$ & $0.159$ & $0.539$ & $0.350$\\
  \hline
 $3$ & $3$ & $0.134$ & $0.821$ & $0.451$\\ 
  \hline
 $2$ & $4$ & $0.124$ & $0.760$ & $0.432$\\
 \hline 
  $3$ & $4$ & $0.437$ & $0.915$ & $0.478$\\ 
  \hline
  $4$ & $4$ & $0.096$ & $1.097$ & $0.523$\\ 
  \hline 
 $2$ & $5$ & $0.102$ & $0.965$ & $0.491$\\ 
  \hline 
 $3$ & $5$ & $0.086$ & $1.072$ & $0.517$\\ 
  \hline 
\end{tabular}
\caption{\label{tab:table-name} Computation of the characteristics time ($T_{-}^{(n), X}$) and the normalized measure of \texttt{SWITCH}-induced memory for different dimensions of the \texttt{SWITCH} control and for different number of channels for $n \le X$, taking $\gamma=2$}\label{table2}
\end{table}

One may also analyse the maximum amount of induced non-Markovian memory that can be efficiently extracted and accessed from the effective action of the cyclic $n$-\texttt{SWITCH} from Fig.~\ref{fig4}. The quantifier tends to saturate with $n$, since the rate of increment decreases as $n$ grows. A trivial upper bound for the induced non-Markovian memory can also be argued from the fact as described below. 

The quantifier for arbitrary $n$ with $n \rightarrow \infty$ becomes
\begin{align}
    &\lim_{n \rightarrow \infty} [\frac{3}{2}(\ln{C^{(n)}}(\infty) - \ln{C^{(n)}}(T_{-}^{(n)})] \nonumber \\
    &~~~~~\overset{a}{=} -\frac{3}{2} \lim_{n \rightarrow \infty} \ln{C^{(n)}} (T_{-}^{(n)}) \overset{b}{\leq} -\frac{3}{2} \ln{C^{(15)}} (T_{-}^{(15)}) = 2.26 \nonumber
\end{align}
The equality (a) arises from the fact that $\lim_{n \rightarrow \infty} [\ln{C^{(n)}}(\infty)] = 0$ (directly derived from the expression of $C^{(n)}$), and the inequality (b) follows since the second term,  $-\ln{C^{(n)}}(T_{-}^{(n)})$, reduces with $n$ and can be trivially bounded for a specific value of $n$, which we set to be $15$ (the bounds become more stringent for higher $n$). Consequently, the maximum value of $M$ as $n \rightarrow \infty$ can be bounded by $2.26$ which in the normalized form becomes $0.693$.

\subsection{Comparison of full-\texttt{SWITCH} consisting of $n!$ dimensional control and cyclic $n$-\texttt{SWITCH} when the number of channels used is fixed}\label{s3e}
To firmly establish the fact that the enhancement of the \texttt{SWITCH}-induced memory actually hinges on the Hilbert-space dimension of the control system, we provide further study on the behavior of the \texttt{SWITCH}-induced memory with varying dimensions ($n$) of the control system while keeping the number of channels ($X$) fixed (however here $n>X$). In particular, we compare the cyclic $n$-\texttt{SWITCH} with $n$ dimensional control with the full \texttt{SWITCH} consisting of $n!$ dimensional control while the number of channels used is fixed to be $n$. 

Since for $n=2$, the full \texttt{SWITCH} as well as the cyclic \texttt{SWITCH} configuration both have a $2$-dimensional control, the first non-trivial example arises for the $n=3$ case. Our primary goal being to compare the effect of the control system's dimension on the memory induced by the quantum \texttt{SWITCH}, we also fix the number of channels to be three i.e., $X=3$. For $X=3$, there are a total of $3!$ possible orderings of the channels, and each configuration can be controlled by a different control system. The Kraus operator corresponding to the superchannel is then described as follows:
    \begin{equation}
    \begin{split}
    S^{X=3}_{i_{1}i_{2}i_{3}} =  &K_{i_{3}}K_{i_{2}}K_{i_{1}} \otimes \ket{0}\bra{0} + K_{i_{1}} K_{i_{3}}K_{i_{2}} \otimes \ket{1}\bra{1}  \\ & + K_{i_{2}}K_{i_{1}}K_{i_{3}}\otimes \ket{2}\bra{2} + K_{i_{2}}K_{i_{3}}K_{i_{1}}\otimes \ket{3}\bra{3} \\ &+ K_{i_{3}} K_{i_{1}}K_{i_{2}}\otimes \ket{4}\bra{4}+ K_{i_{1}}K_{i_{2}}K_{i_{3}}\otimes \ket{5}\bra{5}
    \end{split}
    \end{equation}
    where $\{K_{i_{1}}\}, \{K_{i_{2}}\}, \{K_{i_{3}}\}$ are the Kraus operators of $\mathcal{N}_{1}, \mathcal{N}_{2}$ and $\mathcal{N}_{3}$ respectively. As discussed in section \ref{s2}, we prepare the control system in a superposition state and finally perform measurement on the control system on a superposition basis. For a six dimensional control system the measurement basis can be a Fourier basis with six orthogonal basis elements. In order to analyse the performance of quantum \texttt{SWITCH}, we have to evaluate the normalized measure of the \texttt{SWITCH}-induced memory over all such possible initial and post-selected states to obtain the maximum possible value of the quantifier. It is important to note that, unlike the previous cases, the maximum value of the measure is obtained for specific choices of control state preparation and post-selection corresponding to the following measurement basis element.
    When the control system is initialized in $\ket{+}$ state (where $\ket{+}= \frac{1}{\sqrt{6}}\begin{pmatrix}
1 & 1 & 1 & 1 & 1 & 1 
\end{pmatrix}^T$,'$T$' denotes transpose) and measured in the Fourier basis, the effective target state corresponding to the post-selection in $\ket{-}$ state (where $\ket{-}= \frac{1}{\sqrt{6}}\begin{pmatrix}
1 & 1 & 1 & -1 & -1 & -1 
\end{pmatrix}^T$) becomes $\rho^{(6)}_{X=3}(t) = \Lambda ^{S^{(6)}}_{X=3}(\rho(0))=\langle -|S^{X=3}_{i_{1}i_{2}i_{3}}(\rho(0)\otimes \ket{+}\bra{+})S^{\dagger^{X=3}}_{i_{1}i_{2}i_{3}}|-\rangle$. After explicit calculations, we find the Lindblad coefficient to be
$$\Gamma^{(6)}_{X=3}(\eta)=\frac{3\gamma\eta}{3\eta-2}.$$
The characteristic time turns out to be
\begin{equation}
T_{-}^{(6),X=3} = \frac{-1}{4 \gamma} \ln({\frac{2}{3}}) \nonumber
\end{equation}
 and at $t=T_{-}^{(6),X=3}$, $\Gamma^{(6)}_{X=3}(t) = \infty$. This implies the \texttt{SWITCH}-induced memory,
\begin{equation}
\mathcal{M}_{X=3}^{(6)} =\frac{3}{2} [\ln C^{(6)}_{X=3}(\infty)-\ln C^{(6)}_{X=3}(T_{-}^{(6),X=3})] = \infty
\end{equation} 
However, one may evaluate the normalized measure $\frac{\mathcal{M}_{X=3}^{(6)}}{1+\mathcal{M}_{X=3}^{(6)}}$ which turns out to be $1$. The BLP measure for this case is presented in appendix \ref{G} for convenience.

On the other hand, the normalized measure for the \texttt{SWITCH}-induced memory for the case $n=3, X=3$, (see section \ref{s3b} for details) turns out to be $0.451$. This establishes the fact that the Hilbert-space dimension of the control systems plays an important role in the non-Markovian memory induced by the quantum \texttt{SWITCH}.
\section{Conclusions}\label{s5}
Despite the extensive range of potential applications of quantum \texttt{SWITCH} in information processing tasks, their comprehensive characterization and a complete understanding of the underlying fundamental resources remain elusive even today. In addition to the coherent superposition of alternative configurations of the quantum components, recent studies have emphasized the significance of maintaining coherence in the control system during both preparation and measurement to ensure the advantages associated with the operation of a quantum \texttt{SWITCH} \cite{anand2023emergent}. In this study we investigate how the Hilbert-space dimension of the control system plays a crucial role in ensuring the promising advantages associated with the operation of quantum \texttt{SWITCH}. To be more specific, we focus on quantifying the emergence of non-Markovian memory within the dynamics of the quantum \texttt{SWITCH} and observe that augmenting the Hilbert-space dimension of the control system is linked to an improved performance of the quantum \texttt{SWITCH}, as evidenced by the non-Markovian memory it produces. The interplay between the Hilbert-space dimension of the control system and induced non-Markovianity is a direct evidence that it acts as a resource behind the action of the quantum \texttt{SWITCH}. 

Our study uncovers several potential open problems and avenues for future research. For instance, if it is possible to set an upper limit on the memory induced by the \texttt{SWITCH} for each value of $n$ under some restricted scenario (e.g., restricting to a particular type of channels etc.), then our formalism can be used to witness the dimension of the control system. We keep this aspect open for future investigation. Additionally, one can further explore the role of the Hilbert-space dimension of the control system explicitly on other information processing tasks e.g., the transmission of classical, quantum, or private information through the quantum channel.

\section{Acknowledgements}\label{s6}
The authors acknowledge the anonymous reviewers for their useful comments and suggestions. S.M. and B.M. acknowledge insightful discussions with Sahil Gopalkrishna Naik of SNBNCBS, Kolkata. B.M. acknowledges the DST INSPIRE fellowship program for financial support.

\bibliography{switch}
\appendix
\section{Kraus representation of quantum $2$-\texttt{SWITCH}} \label{A}
For the $2$-\texttt{SWITCH}, the Kraus operator corresponding to the superchannel is described as $$S_{i_{1}i_{2}}=K_{i_{2}}K_{i_{1}} \otimes \ket{0}\bra{0}+K_{i_{1}}K_{i_{2}} \otimes \ket{1}\bra{1}.$$ 
Therefore, the overall evolution of the composite target-control state can be expressed as
\begin{equation}\label{app_W}
    W(\mathcal{N}_1, \mathcal{N}_{2})(\rho \otimes \omega) = \sum_{i_{1}i_{2}} S_{i_{1}i_{2}}(\rho \otimes \omega) S_{i_{1}i_{2}}^{\dagger}.
\end{equation} 
After measuring the control in the coherent basis, the reduced target state of the system corresponding to the $'+'$ outcome becomes,
$$\Lambda^{S^{(2)}}(\rho)=\bra{+}W(\mathcal{N}_1, \mathcal{N}_{2})(\rho \otimes \omega)\ket{+}.$$
Let us now rewrite the action of the quantum $2$-\texttt{SWITCH} in terms of Kraus representation. Eq.\eqref{app_W} can also be written as
\begin{equation}
    \begin{split}
& W(\mathcal{N}_1, \mathcal{N}_{2})(\rho \otimes \ket{+}\bra{+}) = \sum_{i_{1}i_{2}} (K_{i_{2}} K_{i_{1}} \rho K_{i_{1}}^{\dagger}K_{i_{2}}^{\dagger} \otimes \frac{1}{2} \ket{0}\bra{0}\\ 
&~~~~~~ + K_{i_{2}} K_{i_{1}} \rho K_{i_{2}}^{\dagger}K_{i_{1}}^{\dagger} \otimes \frac{1}{2} \ket{0}\bra{1} + K_{i_{1}} K_{i_{2}} \rho K_{i_{1}}^{\dagger}K_{i_{2}}^{\dagger} \otimes \frac{1}{2} \ket{1}\bra{0} \\ 
&~~~~~~ + K_{i_{1}} K_{i_{2}} \rho K_{i_{2}}^{\dagger}K_{i_{1}}^{\dagger} \otimes \frac{1}{2} \ket{1}\bra{1}). \nonumber
\end{split}
\end{equation}
After measuring the control state in the coherent basis the reduced state of target corresponding to $`+'$ outcome becomes (in terms of Kraus operator),
\begin{equation}
    \begin{split}
& \bra{+}W(\mathcal{N}_1, \mathcal{N}_{2})(\rho \otimes \ket{+}\bra{+})\ket{+} =\\ 
& \frac{1}{4}\sum_{i_{1}i_{2}} (K_{i_{2}}K_{i_{1}}\rho K_{i_{1}}^{\dagger} K_{i_{2}}^{\dagger}+K_{i_{2}}K_{i_{1}}\rho K_{i_{2}}^{\dagger} K_{i_{1}}^{\dagger}\\ &+K_{i_{1}}K_{i_{2}}\rho K_{i_{1}}^{\dagger} K_{i_{2}}^{\dagger}+K_{i_{1}}K_{i_{2}}\rho K_{i_{2}}^{\dagger} K_{i_{1}}^{\dagger})
\end{split}
\end{equation}
If we express $\bra{+}W(\mathcal{N}_1, \mathcal{N}_{2})(\rho \otimes \ket{+}\bra{+})\ket{+} = \sum_{i_{1}i_{2}} M_{i_{1}i_{2}} \rho M_{i_{1}i_{2}}^{\dagger}$ then  $M_{i_{1}i_{2}}=\frac{K_{i_{1}}K_{i_{2}}+K_{i_{2}}K_{i_{1}}}{2}$. The normalized target state after the \texttt{SWITCH} action is given by $$\frac{\sum_{i_{1}i_{2}} M_{i_{1}i_{2}} \rho M_{i_{1}i_{2}}^{\dagger}}{\Tr(\sum_{i_{1}i_{2}} M_{i_{1}i_{2}} \rho M_{i_{1}i_{2}}^{\dagger})}$$ 
which is Eq.\eqref{switchkraus}. 

Above analysis can straightforwardly be extended to $n$-\texttt{SWITCH} in a similar fashion.

\section{Derivation of the master equation} \label{B}
Here, we find the Lindblad type master equation of a central spin model \citep{bhattacharya2017exact, mukhopadhyay2017dynamics}. It consists of a qubit system that interacts with a bath $(B)$ comprising of $N$ qubits which can also be called as environment. The system-environment Hamiltonian is given by 
\begin{equation}
    H= \frac{\hbar}{2} \omega_{0}\sigma_{z}^{0} + \frac{\hbar}{2} \sum_{i=1}^{N} b(\sigma_{x}^{0}\sigma_{x}^{i}+\sigma_{y}^{0}\sigma_{y}^{i}+\sigma_{z}^{0}\sigma_{z}^{i})+H_{B}.
\end{equation}
where $\sigma_{k}^{0}$ and  $\sigma_{k}^{i}$ are the Pauli matrices for the system and $i$th spin of bath respectively,  with $k \in \{x,y,z\}$, $b$ is the interaction strength and $H_{B}$ is the bath Hamiltonian. Initially, the system is uncorrelated with the environment and the system-environment state is $\rho(0)\otimes \frac{\mathbb{1}}{2^{N}}$.
The system state $(\rho)$ dynamically evolves as 
\begin{equation}
    \rho (t)= \Tilde{\Phi} (\rho(0))
\end{equation}
where $ \Tilde{\Phi}(.)$ is the linear dynamical map.
The equation of motion of reduced state of the system is 
\begin{equation}
    \dot{\rho}(t)=\Tilde{\Lambda}(\rho(t)) \label{dynamics}
\end{equation}
where $\Tilde{\Lambda}(.)$ is the time-dependent generator of the dynamics. To be a valid master equation, $\Tilde{\Lambda}(\rho)$ must be Hermitian, traceless for all $\rho$ and $\Tilde{\Lambda}$ must be a linear map. Solving the master equation necessarily requires finding the generator. We resort to the methods used in Refs.\citep{andersson2007finding,hall2014canonical,bhattacharya2017exact} in order to find this generator.
$\Tilde{\Phi}(\rho(0))$ can be expressed in terms of orthonormal basis $\{F_{i}\}$ of $\mathbb{C}^{2}$ as 
    \begin{equation}
        \rho(t)=\Tilde{\Phi}(\rho(0))= \sum_{ij}\Tr[\rho(0)F_{i}]\Tr[(\Tilde{\Phi} F_{i})F_{j}]F_{j}  \label{phimap}
         \end{equation}
with the properties $F_{i}^{\dagger} = F_{i}, F_{0}=\frac{\mathbb{1}}{\sqrt{2}}, Tr[F_{i}F_{j}]=\delta_{ij}$ and $F_{i}'$s are traceless operators except $F_{0}$. Eq.\eqref{phimap} then becomes
$$\rho(t)=\sum_{ij}r_{i}(0)G_{ij}F_{j}$$
with $G_{ij}=\Tr[(\Tilde{\Phi} F_{i})F_{j}]$ and $r_{i}(0)=\Tr[\rho(0)F_{i}]$.
The corresponding matrix form is 
   $$ \rho(t)=[G(t)r(0)]^{T} F $$ 
The time-derivative of $\rho(t)$ is,
\begin{equation}
    \dot{\rho}(t)=[\dot{G}(t)r(0)]^{T} F \label{derivative}
\end{equation}
and
\begin{equation}
\Tilde{\Lambda}(\rho(t)) =\sum_{ij}\Tr[\rho(t)F_{i}]\Tr[(\Tilde{\Lambda} F_{i})F_{j}]F_{j}= [L(t)r(t)]^{T} F \label{Lindblad}
\end{equation}
where $L_{ij}=\Tr[(\Tilde{\Lambda} F_{i})F_{j}]$ and $r_{i}(t)=\Tr[\rho(t)F_{i}]$. $L(t)$ and $r(t)$ are the matrix forms of $L_{ij}$ and $r_{i}(t)$ respectively.
Using Eqs.\eqref{dynamics}, \eqref{derivative} and \eqref{Lindblad},
$$\dot{G}(t)r(0)=L(t)r(t)=L(t)G(t)r(0)$$
$$\implies L(t)=\dot{G}(t)G(t)^{-1}.$$
So if the map $G(t)$ is invertible, then it is possible to find $L(t)$, which is indeed the case here. Following the prescription given in Refs.\citep{hall2014canonical,andersson2007finding,bhattacharya2017exact}, the dynamical map of Eq.\eqref{updatedswitchstate} has the form 
\begin{equation}
\begin{split}
  & \rho_{11}(t)= (\frac{1+C^{(n)}(t)}{2}) \rho_{11}(0) + (\frac{1-C^{(n)}(t)}{2}) \rho_{22}(0) \\ &
  \rho_{22}(t)= (\frac{1-C^{(n)}(t)}{2}) \rho_{11}(0) + (\frac{1+C^{(n)}(t)}{2}) \rho_{22}(0) \\ &
  \rho_{12}(t)= C^{(n)}(t) \rho_{12}(0) \\ &
  \rho_{21}(t)= C^{(n)}(t) \rho_{21}(0) = C^{(n)}(t) \rho^{*}_{12}(0)
   \end{split}
\end{equation}
where $C^{(n)}(t)$ is a real-valued function of time. The matrix $L(t)$ is given by
\begin{equation}
 \begin{pmatrix}
  0 & 0 & 0 & 0 \\
0 & \frac{d}{dt} [\ln C^{(n)}(t)] & 0 & 0\\ 
0 & 0 & \frac{d}{dt} [\ln C^{(n)}(t)] & 0\\ 
0 & 0 & 0 & \frac{d}{dt} [\ln C^{(n)}(t)]\\ 
  \end{pmatrix}
\end{equation}
The master equation is 
\begin{equation}
    \frac{d\rho(t)}{dt}= \sum_{i=1}^{3}\Gamma^{(n)}_{i}(t) [\sigma_{i}\rho(t)\sigma_{i}-\rho(t)]
\end{equation}
where $\Gamma^{(n)}_{i}(t)=-\frac{1}{4}\frac{d}{dt} \ln{C^{(n)}(t)} \forall i \in \{1,2,3\}$ .

\section{Linearity of the $n$-\texttt{SWITCH} map} \label{C}
\noindent{\textbf{Proposition 1:}} {\textit{The map under the action of $n$-\texttt{SWITCH} is linear.}}
\begin{proof}
A map $\Lambda$ is said to be linear if for any two states $\rho$ and $\sigma$, 
$$\Lambda(a\rho+b\sigma)=a\Lambda{(\rho)}+b\Lambda{(\sigma)}$$ for arbitrary elements $a$ and $b$. To prove that the $n$-\texttt{SWITCH} map is linear, consider two normalized states $\rho$ and $\sigma$ where $\rho = \begin{bmatrix}
\rho_{11} & \rho_{12} \\
\rho_{21} & \rho_{22} \\
\end{bmatrix}$ and $\sigma = \begin{bmatrix}
\sigma_{11} & \sigma_{12} \\
\sigma_{21} & \sigma_{22} \\
\end{bmatrix}$.
\begin{equation}
\begin{split}
&\Lambda(a\rho+b\sigma)=\Lambda\begin{bmatrix}
a\rho_{11}+b\sigma_{11} & a\rho_{12}+b\sigma_{12} \\
a\rho_{21}+b\sigma_{21}  & a\rho_{22}+b\sigma_{22}\\
\end{bmatrix}
 =\begin{bmatrix}
\alpha & \beta \\
\gamma  & \delta  \label{linearity}
\end{bmatrix}
\end{split}
\end{equation}
where $$\alpha=A^{(n)}(t)(a\rho_{11}+b\sigma_{11})+B^{(n)}(t)(a\rho_{22}+b\sigma_{22}),$$ $$\beta=C^{(n)}(t)(a\rho_{12}+b\sigma_{12}),$$ $$\gamma=C^{(n)}(t)(a\rho_{21}+b\sigma_{21})$$ and $$\delta=B^{(n)}(t)(a\rho_{11}+b\sigma_{11})+A^{(n)}(t)(a\rho_{22}+b\sigma_{22})$$
and the map action is described in Eq.\eqref{updatedswitchstate}. Since $A^{(n)}(t), B^{(n)}(t)$ and $C^{(n)}(t)$ are independent of the states $\rho$ and $\sigma$, the matrix elements of Eq.\eqref{linearity} becomes
$$\alpha=a(A^{(n)}(t)\rho_{11}+B^{(n)}(t)\rho_{22})+b(A^{(n)}(t)\sigma_{11}+B^{(n)}(t)\sigma_{22}),$$
$$\beta=aC^{(n)}(t)\rho_{12}+bC^{(n)}(t)\sigma_{12},$$
$$\gamma=a C^{(n)}(t)\rho_{21} +b  C^{(n)}(t)\sigma_{21}$$ and
$$\delta=a(B^{(n)}(t)\rho_{11}+A^{(n)}(t)\rho_{22})+b(B^{(n)}(t)\sigma_{11}+A^{(n)}(t)\sigma_{22}).$$ This can be written as 
$$a\begin{bmatrix}
A^{(n)}(t)\rho_{11}+B^{(n)}(t)\rho_{22} &  C^{(n)}(t)\rho_{12} \\
  C_{(n)}(t)\rho_{21}  & B^{(n)}(t)\rho_{11}+A^{(n)}(t)\rho_{22} \\
\end{bmatrix}$$  
$$+ b \begin{bmatrix}
A^{(n)}(t)\sigma_{11}+B^{(n)}(t)\sigma_{22} &  C^{(n)}(t)\sigma_{12} \\
  C_{(n)}(t)\sigma_{21}  & B^{(n)}(t)\sigma_{11}+A^{(n)}(t)\sigma_{22} \\
\end{bmatrix}$$
$$= a\Lambda(\rho)+b \Lambda(\sigma).$$ So, the map is linear.
\end{proof}

\section{Action of the quantum 2-\texttt{SWITCH} for $\gamma_{1}(t) \neq \gamma_{2}(t) \neq \gamma_{3}(t)$ }\label{D} 
For simplicity and without loss of generality, we assume that the Lindblad coefficients are all constants (i.e., no time dependence), and hence $\gamma_{i} (t) = \gamma_{i} ~\forall i \in \{1,2,3\}$ and $\forall t$.
For a completely depolarizing channel, the master equation is given by Eq.\eqref{master}. The corresponding Kraus operators are given by Eq.\eqref{kraus}
with 
\begin{equation}
    \begin{split}
    &    \zeta_{1}( t) =  (\gamma_1  + \gamma _2) t \\
     &    \zeta_{2}( t) =  (\gamma_2  + \gamma _3) t. 
     \end{split}
\end{equation}
The state after the channel action is represented by Eq.\eqref{state}. 
 If the initial target qubit state of the system is $\rho(0)$, then the state of the target system after the action of quantum $2$-\texttt{SWITCH} is given by Eq.\eqref{updatedswitchstate}.
 Following a similar prescription as in section \ref{s3}, one  can clearly see that $A^{(2)}(t), B^{(2)}(t)$ and $C^{(2)}(t)$ are functions of $\gamma_{1}, \gamma_{2}, \gamma_{3}$ and $t$, with  $A^{(2)}(t)+B^{(2)}(t)=1$.\\
 Note that unlike the previous case, here $A^{(2)}(t)-B^{(2)}(t) \neq C^{(2)}(t)$.
 The master equation corresponding to the \texttt{SWITCH} dynamical map can be derived in a similar fashion, and is given by Eq.\eqref{switchmaster} with $n=2$. The Lindblad coefficients after the $2$-\texttt{SWITCH} action is given by
$$\Gamma_{1}^{(2)}(t)=\Gamma_{2}^{(2)}(t)=\Gamma^{(2)}(t)=-\frac{1}{4}\frac{d}{dt} \ln{(A^{(2)}(t)-B^{(2)}(t))}$$ and
$$\Gamma_{3}^{(2)}(t)= \frac{1}{4}\frac{d}{dt} \ln{(A^{(2)}(t)-B^{(2)}(t))}-\frac{1}{2}\frac{d}{dt} \ln{(C^{(2)}(t))}.$$ The RHP measure can be easily calculated and it comes out to be $\int_{t=T_{-}}^\infty -2(2 \Gamma^{(2)}(t)+\Gamma_{3}^{(2)}(t)) dt$.

\begin{figure}[ht]
\includegraphics[width=.45\textwidth]{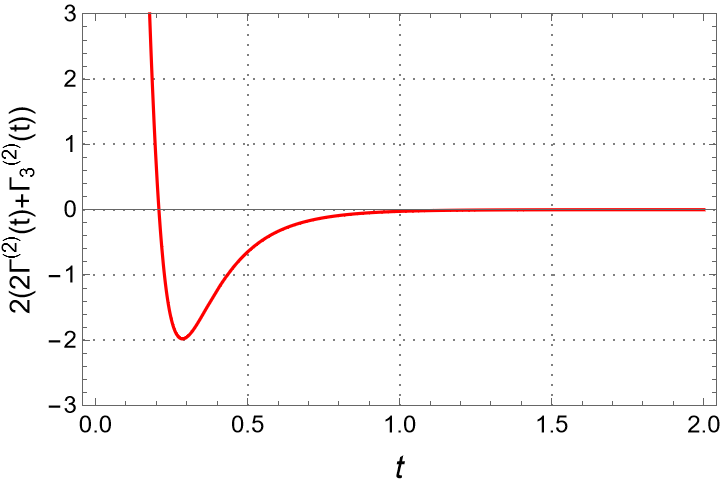} 
\caption{Representation of the \texttt{SWITCH}-induced measure of non-Markovianity in RHP sense for $2$-\texttt{SWITCH} in the region where $2(2\Gamma^{(2)}(t)+\Gamma_{3}^{(2)}(t)) <0$ and $\gamma_{1}=1, \gamma_{2}=2, \gamma_{3}=3$. The negative region indicates the emergent non-Markovianity due to the \texttt{SWITCH} action.}
\centering \label{figg}
\end{figure}
Fig. \ref{figg} shows that there is a \texttt{SWITCH}-induced memory even when the Lindblad coefficients before the \texttt{SWITCH} action aren't considered to be equal. This can be extended in a similar fashion for different dimensions of the control and there will be a memory induced in those cases as well. This is because, even when the initial Lindblad coefficients are not equal, the dynamics remain (non-Markovian) depolarizing after the \texttt{SWITCH} action. We have already demonstrated in various sections of the paper that such depolarizing dynamics exhibit \texttt{SWITCH}-induced non-Markovianity.

\section{Computation of the quantifier for cyclic $n$-\texttt{SWITCH}}  \label{E}
In order to compute the quantifier explicitly, we first need to calculate the characteristics time i.e., the initial time at which information backflow triggers. This can also be achieved by calculating the time when $\Gamma^{(n)}(t)$ changes sign from positive to negative. Below we explicitly calculate the characteristics time and the induced non-Markovian memory corresponding to the cyclic $n$-\texttt{SWITCH} for several $n$.

$\bullet$ \textbf{For  $2$-\texttt{SWITCH}:}
As discussed earlier, the characteristics time, $T_{-}^{(2)}$, is the time at which $\Gamma^{(2)}(t)=0$ . From the expression
\begin{equation}
        \Gamma^{(2)}(\eta)=\frac{16 \gamma \eta(3 \eta^{2} +6 \eta -1)}{(9 \eta^{2}-2 \eta +1)(-3 \eta^{2}+6\eta+5)} , \nonumber
\end{equation}
this reduces to  $3 \eta^{*2} +6 \eta^{*}-1=0$ where $\eta* = \eta(T_{-}^{(2)})$. Solving this, we get 
\begin{equation}
T_{-}^{(2)} = -\frac{1}{4\gamma} \ln{(0.155)}. \nonumber
\end{equation}

\begin{equation}
   \implies \mathcal{M}^{(2)}=  \frac{3}{2} [\ln C^{(2)}(\infty)-\ln C^{(2)}(T_{-}^{(2)})] =0.385.
\end{equation}
 The normalized measure is $0.278$.

$\bullet$ \textbf{For  $3$-\texttt{SWITCH}:}
For $3$-\texttt{SWITCH}, we have the expression for $\Gamma^{(3)}(\eta)$ as 
\begin{equation}
 \Gamma^{(3)}(\eta)=\frac{-2 \gamma \eta(-3\eta^{4}-6\eta^{3}-12\eta^{2} +2 \eta +1)}{(-7\eta^{6}+8\eta^{5}+7\eta^{4}+4\eta^{3}-\eta^{2}+1)}. \nonumber
\end{equation}

At $t=T_{-}^{(3)}$, $\Gamma^{(3)}(t) = 0$. Therefore the condition further reduces to $3 \eta^{*4} + 6 \eta^{*3} + 12 \eta^{*2} -2 \eta^{*}-1 = 0$ where $\eta* = \eta(T_{-}^{(3)})$. Solving this, we get 
\begin{equation}
T_{-}^{(3)} = -\frac{1}{4\gamma} \ln{(0.342)} \nonumber
\end{equation}

\begin{equation}
   \implies \mathcal{M}^{(3)}=\frac{3}{2} [\ln C^{(3)}(\infty)-\ln C^{(3)}(T_{-}^{(3)})] = 0.821.
\end{equation}
The normalized measure is $0.451$.

$\bullet$ In a similar way, above analysis can be extended to $n$-\texttt{SWITCH} where $n$ can be arbitrarily large. However, below (in Table \ref{table1}) we summarize the results only upto $n=15$.
\begin{table}[h!]
\begin{tabular}{| c| c| c| c |c |c |}
\hline
  Dimension $(n)$ & $T^{(n)}_{-}$  & Measure& Normalized Measure \\
   & &${(\mathcal{M}^{(n)}})$  & $ (\frac{\mathcal{M}^{(n)}}{1+\mathcal{M}^{(n)}})$ \\
 
 \hline 
 $2$  & $0.233$ & $0.385$ & $0.278$\\  
  \hline
 $3$ & $0.134$ & $0.821$ & $0.451$\\
  \hline 
 $4$ & $0.096$ & $1.097$ & $0.523$\\ 
 \hline 
 $5$ & $0.075$ & $1.284$ & $0.562$\\ 
  \hline 
 $6$ & $0.062$ & $1.418$ & $0.586$\\ 
  \hline 
 $7$ & $0.053$ & $1.520$ & $0.603$\\
 \hline 
 $8$ &  $0.046$ & $1.599$ & $0.615$\\
 \hline 
 $9$ & $0.041$ & $1.663$ & $0.624$\\
 \hline 
 $10$ & $0.037$ & $1.715$ &$0.632$ \\
 \hline 
 $11$ & $0.033$ & $1.758$ &$0.637$ \\
 \hline 
 $12$ & $0.030$ & $1.795$ &$0.642$ \\
 \hline 
 $13$ & $0.028$ & $1.827$ &$0.646$ \\
 \hline 
 $14$ & $0.026$ & $1.855$ &$0.650$ \\
 \hline 
 $15$ & $0.024$ & $1.879$ &$0.653$ \\
 \hline 
\end{tabular}
\caption{\label{tab:table-name} Computation of characteristics time and normalized measure for different dimensions of the \texttt{SWITCH} control for $n=X$, taking $\gamma=2$.}\label{table1}
\end{table}

\section{Case study of $n$-\texttt{SWITCH} using $X$ number of channels, where $n \le X$.}  \label{F}

Here, we explicitly calculate the \texttt{SWITCH}-induced memory of an $n$-\texttt{SWITCH} using $X$ channels, where the Hilbert-space dimension of the control system $n$ is less than the number of channels, $X$ used for the quantum \texttt{SWITCH}. It can be readily verified that for the aforementioned cases, the measure of \texttt{SWITCH}-induced memory is maximum when the initial state is chosen to be $\ket{+}$ followed by a post-selection in the diagonal basis corresponding to the $'+'$ outcome, whereas a post-selection corresponding to the $'-'$ outcome ($\ket{-}$ denotes the state(s) orthogonal to $\ket{+}$) gives the value of measure to be $0$. Note that we shall use the superscripts in $S^{X}$, $M^{X}, T_{-}^{X}$ and the subscripts in $\rho_{X}(t), \eta_{X}, C_{X}$, $\Lambda_{X}$, $\mathcal{M}_{X}$ and $\Gamma_{X}$ when $n \ne X$ whereas for $n=X$, i.e., for the cyclic cases (presented in section \ref{s3}), these are dropped for convenience.\\
\begin{itemize}
    \item \textbf{For  $n=2, X=3$:} With a two-dimensional control system, only two possible configurations can be effectively controlled. However, with three channels available, two channels will always have a definite order of action, while the third channel will be in an indefinite causal order relative to the overall configuration of the first two. Without loss of generality, we can assign one control to $\mathcal{N}_{2} \circ \mathcal{N}_{1}$ and the other to $\mathcal{N}_{3}$. If the control is prepared in $\ket{0}\bra{0}$ state, the input state first goes through $\mathcal{N}_{2} \circ \mathcal{N}_{1}$, and then through $\mathcal{N}_{3}$ whereas if the control is prepared in $\ket{1}\bra{1}$ state, the input state first goes through $\mathcal{N}_{3}$ and then through $\mathcal{N}_{2} \circ \mathcal{N}_{1}$. The Kraus operator corresponding to the superchannel can then be described as 
    \begin{equation}
    \begin{split}
    S^{X=3}_{i_{1}i_{2}i_{3}}= & K_{i_{3}}K_{i_{2}}K_{i_{1}} \otimes \ket{0}\bra{0} \\ & + K_{i_{2}}K_{i_{1}}K_{i_{3}} \otimes \ket{1}\bra{1}
    \end{split}
    \end{equation}
    where $K_{i_{1}}$, $K_{i_{2}}$ and $K_{i_{3}}$ are the corresponding Kraus operators of those channels. When the control is initialised in $\ket{+}$ state and finally measured in the $\{\ket{+},\ket{-}\}$ basis, the effective target state (corresponding to the `$+$' outcome) becomes $\rho^{(2)}_{x=3}(t) = \Lambda ^{S^{(2)}}_{X=3}(\rho(0))=\langle +|S^{X=3}_{i_{1}i_{2}i_{3}}(\rho(0)\otimes \ket{+}\bra{+})S^{\dagger^{X=3}}_{i_{1}i_{2}i_{3}}|+\rangle$. $\rho^{(2)}_{X=3}(t)$ can also be written in an alternative form: $$\rho^{(2)}_{X=3}(t)=\frac{\sum_{i_{1}i_{2}} M^{X=3}_{i_{1}i_{2}i_{3}} \rho(0) M_{i_{1}i_{2}i_{3}}^{\dagger^{X=3}}}{\Tr(\sum_{i_{1}i_{2}i_{3}} M^{X=3}_{i_{1}i_{2}i_{3}} \rho(0) M^{\dagger ^{X=3}}_{i_{1}i_{2}i_{3}})}$$
with $M^{X=3}_{i_{1}i_{2}i_{3}} = \frac{K_{i_{3}} K_{i_{2}} K_{i_{1}} +  K_{i_{2}}  K_{i_{1}} K_{i_{3}}}{2}$ and $ i_{1},i_{2}, i_{3}\in \{1,2,3,4\}$. It may be noted here that the form of the final state remains same as Eq. \eqref{updatedswitchstate} and thus we will arrive at the same form of the Master equation. After simplification and a comparison with Eq. \eqref{updatedswitchstate}, we obtain 
$$C^{(2)}_{X=3}(\eta)= \frac{-9\eta^{3}+\eta^{2}+\eta-1}{3\eta^{3}-3\eta^{2}-3\eta-5}.$$ This leads to the Lindblad coefficient,
\begin{equation}
 \begin{split}   
\hspace{0.5cm} &\Gamma^{(2)}_{X=3}(\eta)\\
 &=\frac{-8\gamma\eta(-3\eta^{4}-6\eta^{3}-18\eta^{2}+2\eta+1)}{(9\eta^{3}-\eta^{2}-\eta+1)(-3\eta^{3}+3\eta^{2}+3\eta+5)}
\end{split}
\end{equation}
At $t=T_{-}^{(2),X=3}$, $\Gamma^{(2)}_{X=3}(t) = 0$. The condition further reduces to $3 \eta^{*4} + 6 \eta^{*3} + 18 \eta^{*2} -2 \eta^{*}-1 = 0$ where $\eta^{*} = \eta(T_{-}^{(2),X=3})$. Solving this, we get the characteristics time 
\begin{equation}
T_{-}^{(2),X=3} = -\frac{1}{4\gamma} \ln{(0.280)} \nonumber
\end{equation}
This implies the measure of the \texttt{SWITCH}-induced memory:
\begin{equation}
\begin{split}
 \hspace{0.5cm}&\mathcal{M}_{X=3}^{(2)}\\
 &= \frac{3}{2} [\ln C^{(2)}_{X=3}(\infty)-\ln C^{(2)}_{X=3}(T_{-}^{(2),X=3})] = 0.539
 \end{split}
\end{equation}
The normalized measure is $\frac{0.539}{0.539+1}=0.350$

In principle, one can also post-select the target state corresponding the $'-'$ outcome (where $\ket{-}=\frac{\ket{0}-\ket{1}}{\sqrt{2}}$. However, this will gives the value of RHP measure to be $0$. So, in order to evaluate the performance of quantum \texttt{SWITCH}, we need to consider the optimum value of the normalized measure, i.e., $\text{max} \{0,0.350\}=0.350$.

\item \textbf{For  $n=2, X=4$:} Since the dimension of the control system ($n=2$) is less than the the number of channels used ($X=4$), here again effective only two possible configurations are possible-- (i) one channel to be controlled by one control system and a combined three channels to be controlled by the other control system., (ii) combined two channels to be controlled by one control system and the rest two channels (combined) by the other control system. 

Let us consider the first case. Without loss of generality, if the control is prepared in $\ket{0}\bra{0} (\ket{1}\bra{1})$, the input state first goes through $\mathcal{N}_{3} \circ \mathcal{N}_{2} \circ \mathcal{N}_{1}(\mathcal{N}_{4})$, and then through $\mathcal{N}_{4}  (\mathcal{N}_{3} \circ \mathcal{N}_{2} \circ \mathcal{N}_{1})$. The Kraus operator corresponding to the superchannel is then described as 
    \begin{equation}
    \begin{split}
    S^{X=4}_{i_{1}i_{2}i_{3}i_{4}}= & K_{i_{4}}K_{i_{3}}K_{i_{2}}K_{i_{1}} \otimes \ket{0}\bra{0} \\ & + K_{i_{3}}K_{i_{2}}K_{i_{1}}K_{i_{4}} \otimes \ket{1}\bra{1}
    \end{split}
    \end{equation}
    where $\{K_{i_{3}}K_{i_{2}}K_{i_{1}}\}$ is the effective Kraus operator of $\mathcal{N}_{3} \circ \mathcal{N}_{2} \circ \mathcal{N}_{1} $ and $\{K_{i_{4}}\}$ is the Kraus operator of $\mathcal{N}_{4}$. When the control is initialized in $\ket{+}$ state and measured in the $\{\ket{+},\ket{-}\}$ basis, the effective target state (corresponding to the `$+$' outcome) becomes $\rho^{(2)}_{x=4}(t) = \Lambda ^{S^{(2)}}_{X=4}(\rho(0))=\langle +|S^{X=4}_{i_{1}i_{2}i_{3}i_{4}}(\rho(0)\otimes \ket{+}\bra{+})S^{\dagger^{X=4}}_{i_{1}i_{2}i_{3}i_{4}}|+\rangle$. $\rho^{(2)}_{X=4}(t)$ can also be written in an alternative form as $$\rho^{(2)}_{X=4}(t)=\frac{\sum_{i_{1}i_{2}i_{3}i_{4}} M^{X=4}_{i_{1}i_{2}i_{3}i_{4}} \rho(0) M_{i_{1}i_{2}i_{3}i_{4}}^{\dagger^{X=4}}}{\Tr(\sum_{i_{1}i_{2}i_{3}i_{4}} M^{X=4}_{i_{1}i_{2}i_{3}i_{4}} \rho(0) M^{\dagger ^{X=4}}_{i_{1}i_{2}i_{3}i_{4}})}$$
with $M^{X=4}_{i_{1}i_{2}i_{3}i_{4}} = \frac{K_{i_{4}}K_{i_{3}} K_{i_{2}} K_{i_{1}} +  K_{i_{3}}  K_{i_{2}}K_{i_{1}} K_{i_{4}}}{2}$ and $ i_{1},i_{2}, i_{3}, i_{4}\in \{1,2,3,4\}$. Since the form of the final state remains same, after simplification and a comparison with Eq.\eqref{updatedswitchstate}, we obtain 
$$C^{(2)}_{X=4}(\eta)= \frac{9\eta^{4}-\eta^{3}-\eta+1}{-3\eta^{4}+3\eta^{3}+3\eta+5}.$$  
Following similar tools as before, we obtain the expression for the characteristics time:
\begin{equation}
T_{-}^{(2),X=4} = -\frac{1}{4\gamma} \ln{(0.372)} \nonumber
\end{equation}
This leads to the measure of the \texttt{SWITCH}-induced memory:
\begin{equation}
\begin{split}
 \hspace{0.5cm}&\mathcal{M}_{X=4}^{(2)} \\
&=\frac{3}{2} [\ln C^{(2)}_{X=4}(\infty)-\ln C^{(2)}_{X=4}(T_{-}^{(2),X=4})] = 0.760
\end{split}
\end{equation}
The normalized measure can be calculated in a similar way, as discussed previously and it turns out to be $0.432$.

For the second case, without loss of generality, if the control is prepared in $\ket{0}\bra{0} (\ket{1}\bra{1})$, the input state first goes through $\mathcal{N}_{2} \circ \mathcal{N}_{1}(\mathcal{N}_{4} \circ \mathcal{N}_{3})$, and then through $\mathcal{N}_{4} \circ \mathcal{N}_{3} (\mathcal{N}_{2} \circ \mathcal{N}_{1})$. The Kraus operator corresponding to the superchannel is then described as 
    \begin{equation}
    \begin{split}
   \hspace{0.5cm} S^{X=4}_{i_{1}i_{2}i_{3}i_{4}}
    = & K_{i_{4}}K_{i_{3}}K_{i_{2}}K_{i_{1}} \otimes \ket{0}\bra{0} \\ & + K_{i_{2}}K_{i_{1}}K_{i_{4}}K_{i_{3}} \otimes \ket{1}\bra{1}
    \end{split}
    \end{equation}
    where $\{K_{i_{2}}K_{i_{1}}\}$ is the effective Kraus operator of $\mathcal{N}_{2} \circ \mathcal{N}_{1}$ and $\{K_{i_{4}}K_{i_{3}}\}$ is the effective Kraus operator of $\mathcal{N}_{4} \circ \mathcal{N}_{3}$. Following similar tools as discussed in the first case, the characteristics time comes out to be:
\begin{equation}
T_{-}^{(2),X=4} = -\frac{1}{4\gamma} \ln{(0.393)} \nonumber
\end{equation}
Therefore, the measure of the \texttt{SWITCH}-induced memory is :
\begin{equation}
\begin{split}
\hspace{0.5cm}&\mathcal{M}_{X=4}^{(2)}\\
&=\frac{3}{2} [\ln C^{(2)}_{X=4}(\infty)-\ln C^{(2)}_{X=4}(T_{-}^{(2),X=4})] = 0.385
\end{split}
\end{equation}
 The normalized measure is $0.278$
Thus the \texttt{SWITCH} quantifier for $n=2, X=4$ is finally obtained by optimizing over these two configurations: $\max \{0.432, 0.278\} = 0.432$

\item \textbf{For $n=3, X=4$ :} In this case, since there are four channels to be controlled with three dimensional control system, one has to use one control system to control two channels combinedly. Without loss of generality we consider, following combination:  $\mathcal{N}_{2} \circ \mathcal{N}_{1}$ to be controlled by one control system and $\mathcal{N}_{3}$, $\mathcal{N}_{4}$ by the other two. For a three-dimensional control, we will have a cyclic superposition of these orders. The Kraus operator corresponding to the superchannel is then described as 
    \begin{equation}
    \begin{split}
 \hspace{0.5cm}  & S^{X=4}_{i_{1}i_{2}i_{3}i_{4}}  =   K_{i_{4}}K_{i_{3}}K_{i_{2}}K_{i_{1}} \otimes \ket{0}\bra{0} \\
 \hspace{0.5cm} & + K_{i_{2}}K_{i_{1}}K_{i_{4}}K_{i_{3}} \otimes \ket{1}\bra{1}  + K_{i_{3}}K_{i_{2}}K_{i_{1}}K_{i_{4}} \otimes \ket{2}\bra{2} 
    \end{split}
    \end{equation}
    where $\{K_{i_{2}}K_{i_{1}}\}$ is the effective Kraus operator of $\mathcal{N}_{2} \circ \mathcal{N}_{1}$, $\{K_{i_{3}}\}$ is the Kraus operator of $\mathcal{N}_{3}$ and $\{K_{i_{4}}\}$ is the Kraus operator of $\mathcal{N}_{4}$. 
 
Using the same procedure as discussed for the earlier case, we evaluate the characteristics time to be $T_{-}^{(3),X=4} = -\frac{1}{4\gamma} \ln{(0.437)}$
and the \texttt{SWITCH}-induced measure
\begin{equation}
\begin{split}
\hspace{0.5cm} &\mathcal{M}_{X=4}^{(3)}\\
 &= \frac{3}{2} [\ln C^{(3)}_{X=4}(\infty)-\ln C^{(3)}_{X=4}(T_{-}^{(3),X=4})] = 0.915
 \end{split}
\end{equation}
The normalized measure is $0.478$.
Note that a post-selection corresponding the $'-'$ outcome (where $\ket{-}$ can be either $\frac{1}{\sqrt{3}} \begin{pmatrix}
1 & \omega ^{2} & \omega 
\end{pmatrix}^T$,$'T'$ denoting transpose or $\frac{1}{\sqrt{3}} \begin{pmatrix}
1 & \omega  & \omega^{2} 
\end{pmatrix}^T$, with $1, \omega, \omega^{2}$ being the cube roots of unity) gives the value of the normalized measure to be $0$. So, the optimized value of the measure in this configuration turns out to be $\max \{0,0.478\}=0.478$
 \item \textbf{For  $n=4, X=4$:} In the case of $4$-\texttt{SWITCH} with $4$ channels, the overall Kraus operator can be expressed as
\begin{equation}
\begin{split}
  \hspace{0.5cm}  S_{i_{1}i_{2}i_{3}i_{4}}
 &  =   K_{i_{4}}K_{i_{3}}K_{i_{2}}K_{i_{1}} \otimes \ket{0}\bra{0}
   +K_{i_{1}}K_{i_{4}}K_{i_{3}}K_{i_{2}} \otimes \ket{1}\bra{1} \\
   & +K_{i_{2}}K_{i_{1}}K_{i_{4}}K_{i_{3}} \otimes \ket{2}\bra{2}+K_{i_{3}}K_{i_{2}}K_{i_{1}}K_{i_{4}} \otimes \ket{3}\bra{3}
    \end{split}
\end{equation}
Following the action of the $4$-\texttt{SWITCH}, when the control qubit is measured in the coherent basis, the resulting target state (associated with the `$+$' outcome) becomes

\begin{equation}
\begin{split}
  & \rho^{(4)}(t)= \Lambda ^{S^{(4)}}(\rho(0))\\
  &=\langle +|S_{i_{1}i_{2}i_{3}i_{4}}(\rho(0)\otimes \ket{+}\bra{+})S_{i_{1}i_{2}i_{3}i_{4}}^{\dagger}|+\rangle \nonumber
   \end{split}
\end{equation}
where $i_{1},i_{2},i_{3},i_{4}\in\{1,2,3,4\}$. Using Eq.\eqref{switchkraus}, we express $\rho^{(4)}(t)$ in an alternative form as 
    $$\rho^{(4)}(t)= \frac{\sum_{i_{1}i_{2}i_{3}i_{4}} M_{i_{1}i_{2}i_{3}i_{4}} \rho(0) M_{i_{1}i_{2}i_{3}i_{4}}^{\dagger}}{\Tr(\sum_{i_{1}i_{2}i_{3}i_{4}} M_{i_{1}i_{2}i_{3}i_{4}} \rho(0) M_{i_{1}i_{2}i_{3}i_{4}}^{\dagger})} $$ 
where $M_{i_{1}i_{2}i_{3}i_{4}}$ =$$
\frac{K_{i_{4}}K_{i_{3}}K_{i_{2}}K_{i_{1}}+ K_{i_{1}}K_{i_{4}}K_{i_{3}}K_{i_{2}}+K_{i_{2}}K_{i_{1}}K_{i_{4}}K_{i_{3}}+K_{i_{3}}K_{i_{2}}K_{i_{1}}K_{i_{4}}}{4}$$
After simplification and a comparison with Eq.\eqref{updatedswitchstate}, we obtain 
  \begin{equation}
 C^{(4)}(\eta) = \frac{-19\eta^{4}+2\eta^{3}+2\eta^{2}+2\eta -3 }{9\eta^{4}-6\eta^{3}-6\eta^{2}-6\eta -7 } \nonumber
 \end{equation}
This suggests that
\begin{align}
\hspace{0.5cm}&\Gamma^{(4)}(\eta)=\nonumber \\
& \frac{-32\gamma\eta(-3\eta^{6}-6\eta^{5}-9\eta^{4}-20\eta^{3}+3\eta^{2}+2\eta+1)}{(19\eta^{4}-2\eta^{3}-2\eta^{2}-2\eta+3)(-9\eta^{4}+6\eta^{3}+6\eta^{2}+6\eta+7)} \label{gamma4}
\end{align}
We now evaluate the characteristics time, $T_{-}^{(4)}$ for the $4$-\texttt{SWITCH}. From Eq.\eqref{gamma4}, it turns out to be $T_{-}^{(4)} = -\frac{1}{4\gamma} \ln{(0.464)} \implies  \mathcal{M}^{(4)}
=\frac{3}{2} [\ln C^{(4)}(\infty)-\ln C^{(4)}(T_{-}^{(4)})]=1.097$. The normalized measure is $\frac{1.097}{1.097+1}=0.523$.
\item \textbf{For  $n=2, X=5$:} Here, again the dimension of the control qubit being two, we can only consider two possible configurations although the number of channels are five. We can combine them in two ways-- (i) By clustering three channels vs two channels. (ii) By clustering four channels vs one channel. 

Consider the first case. Without loss of generality, if the control is $\ket{0}\bra{0} (\ket{1}\bra{1})$, the input state first goes through $\mathcal{N}_{3} \circ \mathcal{N}_{2} \circ \mathcal{N}_{1} (\mathcal{N}_{5} \circ \mathcal{N}_{4})$, and then through $\mathcal{N}_{5} \circ \mathcal{N}_{4} (\mathcal{N}_{3} \circ \mathcal{N}_{2} \circ \mathcal{N}_{1})$. The Kraus operator corresponding to the superchannel is then described as 
    \begin{equation}
    \begin{split}
    S^{X=5}_{i_{1}i_{2}i_{3}i_{4}i_{5}}= & K_{i_{5}}K_{i_{4}}K_{i_{3}}K_{i_{2}}K_{i_{1}} \otimes \ket{0}\bra{0} \\ & + K_{i_{3}}K_{i_{2}}K_{i_{1}}K_{i_{5}}K_{i_{4}} \otimes \ket{1}\bra{1}
    \end{split}
    \end{equation}
    where $\{K_{i_{3}}K_{i_{2}}K_{i_{1}}\}$ is the effective Kraus operator of $\mathcal{N}_{3} \circ \mathcal{N}_{2}\circ \mathcal{N}_{1}$ and $\{K_{i_{5}}K_{i_{4}}\}$ is the effective Kraus operator of $\mathcal{N}_{5} \circ \mathcal{N}_{4}$. 

We proceed in the same way as before and obtain the characteristics time $T_{-}^{(2),X=5} = -\frac{1}{4\gamma} \ln{(0.471)} $, and the \texttt{SWITCH}-induced measure

\begin{equation}
\begin{split}
\hspace{0.5cm} &\mathcal{M}_{X=5}^{(2)}\\
 &=\frac{3}{2} [\ln C^{(2)}_{X=5}(\infty)-\ln C^{(2)}_{X=5}(T_{-}^{(2),X=5})] = 0.438
 \end{split}
\end{equation}
The normalized measure will be $0.304$.

For the second case, if the control is $\ket{0}\bra{0} (\ket{1}\bra{1})$, the input state first goes through $\mathcal{N}_{4} \circ \mathcal{N}_{3} \circ \mathcal{N}_{2} \circ \mathcal{N}_{1} (\mathcal{N}_{5})$, and then through $\mathcal{N}_{5} (\mathcal{N}_{4} \circ \mathcal{N}_{3} \circ \mathcal{N}_{2} \circ \mathcal{N}_{1})$. The Kraus operator corresponding to the superchannel is then described as 
    \begin{equation}
    \begin{split}
    S^{X=5}_{i_{1}i_{2}i_{3}i_{4}i_{5}}= & K_{i_{5}}K_{i_{4}}K_{i_{3}}K_{i_{2}}K_{i_{1}} \otimes \ket{0}\bra{0} \\ & + K_{i_{4}}K_{i_{3}}K_{i_{2}}K_{i_{1}}K_{i_{5}} \otimes \ket{1}\bra{1}
    \end{split}
    \end{equation}
    where $\{K_{i_{4}}K_{i_{3}}K_{i_{2}}K_{i_{1}}\}$ is the effective Kraus operator of $\mathcal{N}_{4} \circ \mathcal{N}_{3}\circ \mathcal{N}_{2}\circ \mathcal{N}_{1}$ and $\{K_{i_{5}}\}$ is the  Kraus operator of $\mathcal{N}_{5}$. By proceeding in the same way as before, the characteristics time $T_{-}^{(2),X=5} = -\frac{1}{4\gamma} \ln{(0.441)} $, and the \texttt{SWITCH}-induced measure turns out to be 
    \begin{equation}
    \begin{split}
\hspace{0.5cm} &\mathcal{M}_{X=5}^{(2)}\\
 &=\frac{3}{2} [\ln C^{(2)}_{X=5}(\infty)-\ln C^{(2)}_{X=5}(T_{-}^{(2),X=5})] = 0.965
 \end{split}
\end{equation}
The normalized measure is $0.491$.
Therefore, the effective normalized measure is $\max \{0.304, 0.491\}=0.491$.

\item \textbf{For  $n=3, X=5$:} We now have five channels to be controlled by a three dimensional control system. Since we can only control effectively three possible configurations, we can consider combined three channels to be controlled by one control system and rest two by two different control systems or combined two channels to be controlled by one control system and another two by other control system and rest one by the remaining control system. 

Consider the first case. Without loss of generality, we consider the following combination. $\mathcal{N}_{3} \circ \mathcal{N}_{2} \circ \mathcal{N}_{1}$ is controlled by one control system, $\mathcal{N}_{4}, \mathcal{N}_{5}$ by the other two. We shall have a cyclic superposition of these orders. The Kraus operator corresponding to the superchannel is then described as 
    \begin{equation}
    \begin{split}
S^{X=5}_{i_{1}i_{2}i_{3}i_{4}i_{5}}= & K_{i_{5}}K_{i_{4}}K_{i_{3}}K_{i_{2}}K_{i_{1}} \otimes \ket{0}\bra{0} \\ & + K_{i_{3}}K_{i_{2}}K_{i_{1}}K_{i_{5}}K_{i_{4}} \otimes \ket{1}\bra{1}  \\ & + K_{i_{4}}K_{i_{3}}K_{i_{2}}K_{i_{1}}K_{i_{5}} \otimes \ket{2}\bra{2}
\end{split}
    \end{equation}
    where $\{K_{i_{3}}K_{i_{2}}K_{i_{1}}\}$ is the effective Kraus operator of $\mathcal{N}_{3} \circ \mathcal{N}_{2}\circ \mathcal{N}_{1}$, $\{K_{i_{4}}\}$ and $\{K_{i_{5}}\}$ are the Kraus operators of $\mathcal{N}_{4}$ and $\mathcal{N}_{5}$ respectively. Note that one can in principle, consider some other ways of combining the channels. But since all the channels are taken to be the same for our purpose, those combinations will trivially fall into one or the other of these categories. Proceeding in a similar way as the other cases, the characteristic time $T_{-}^{(2),X=5} = -\frac{1}{4\gamma} \ln{(0.501)} $, and the \texttt{SWITCH}-induced measure comes out to be 
    \begin{equation}
    \begin{split}
\hspace{0.5cm}& \mathcal{M}_{X=5}^{(3)}\\
&=\frac{3}{2} [\ln C^{(3)}_{X=5}(\infty)-\ln C^{(3)}_{X=5}(T_{-}^{(3),X=5})] = 1.072
 \end{split}
\end{equation}
The normalized measure is $0.517$.
    
Let us now consider the second case. Without loss of generality we take the following combination: $\mathcal{N}_{2} \circ \mathcal{N}_{1}$,  $\mathcal{N}_{4} \circ \mathcal{N}_{3}$ and $\mathcal{N}_{5}$. For a three-dimensional control, we consider the cyclic superposition of these orders. The Kraus operator corresponding to the superchannel is then described as 
    \begin{equation}
    \begin{split}
    S^{X=5}_{i_{1}i_{2}i_{3}i_{4}i_{5}} =  &K_{i_{5}} K_{i_{4}}K_{i_{3}}K_{i_{2}}K_{i_{1}} \otimes \ket{0}\bra{0} \\ & + K_{i_{2}}K_{i_{1}} K_{i_{5}}K_{i_{4}}K_{i_{3}} \otimes \ket{1}\bra{1}  \\ & + K_{i_{4}}K_{i_{3}}K_{i_{2}}K_{i_{1}}K_{i_{5}}\otimes \ket{2}\bra{2} 
    \end{split}
    \end{equation}
  where $\{K_{i_{2}}K_{i_{1}}\}$ is the effective Kraus operator of $\mathcal{N}_{2} \circ \mathcal{N}_{1}$, $\{K_{i_{4}} K_{i_{3}}\} $ is the effective Kraus operator of $\mathcal{N}_{4} \circ \mathcal{N}_{3}$ and $\{K_{i_{5}}\}$ is the Kraus operator of $\mathcal{N}_{5}$. 

After explicit calculation we find the characteristics time, $T_{-}^{(3),X=5} = -\frac{1}{4\gamma} \ln{(0.517)}$ and the \texttt{SWITCH}-induced measure

\begin{equation}
\begin{split}
  \hspace{0.5cm}  &\mathcal{M}_{X=5}^{(3)}\\
    &= \frac{3}{2} [\ln C^{(3)}_{X=5}(\infty)-\ln C^{(3)}_{X=5}(T_{-}^{(3),X=5})] = 0.886
    \end{split}
\end{equation}
The normalized measure is $0.470$.
Thus the effective normalized measure for $n=3, X=5$ is then $\max \{0.470, 0.517\}=0.517.$
\end{itemize}
We summarize our findings in the table \ref{table2}.

\section{Evaluating the BLP measure for the full-\texttt{SWITCH} (of $3$ channels) having $3!$ dimensional control and cyclic $3$-\texttt{SWITCH} having $3$ dimensional control}  \label{G}
Here, we calculate the BLP (Breuer, Laine, and Piilo) measure \citep{breuer2009measure} for the $n = 6, X=3$ (presented in section \ref{s3e}) and $n = 3, X=3$ (presented in section \ref{s3b}) case. BLP measure quantifies the distance between the two states, before and after the action of the dynamical map. Due to interaction with noisy environment, two quantum states tend to lose their distinguishability gradually with time and hence, the distance between them decreases with time. However, if at any instant of time the distinguishability and distance increases, then there is a backflow of information from the environment to the system leading to the signature of non-Markovianity. 
The distance between two quantum states $\rho_{1}$ and $\rho_{2}$ is defined as $\mathcal{D} (\rho_{1},\rho_{2})=||\rho_{1}-\rho_{2}|| = Tr \sqrt{(\rho_{1}-\rho_{2})^{+}(\rho_{1}-\rho_{2})}$. For simplicity and without any loss of generality, we choose the state $\rho_{2}$ to be the fixed point of the dynamics ($\tau$). The fixed point of a dynamics is the one that remains invariant under the action of the dynamical map. For the \texttt{SWITCH} dynamical map, the fixed point is $\frac{\mathbb{1}}{2}$ \cite{anand2023emergent}. The time derivative of the distance between two states $\rho_{1}=\rho$ and $\rho_{2}=\tau$, evolving under the full-\texttt{SWITCH} action (of three channels) is given by 
\begin{equation}
    \mathcal{B}=\frac{d}{dt}\mathcal{D}(\Lambda _{X=3}^{S^{(6)}}(\rho), \tau)
\end{equation}
The BLP measure is defined to be 
\begin{equation}
\begin{split}
   & \mathcal{M}_{\text{BLP}}  =   \underset{\rho} {\mathrm{max}}\int_{T_{-}}^{\infty} \mathcal{B} \,dt \\ & = \underset{\rho} {\mathrm{max}} [\mathcal{D}(\Lambda _{X=3}^{S^{(6)}}(\rho), \tau)\vert_{t=\infty}-\mathcal{D}(\Lambda _{X=3}^{S^{(6)}}(\rho), \tau)\vert_{t=T_{-}^{(6),X=3}}]
\end{split}
\end{equation}
The normalized BLP measure is given by $\frac{\mathcal{M}_{\text{BLP}}}{\mathcal{M}_{\text{BLP}}+1}$.
After simplification, this reduces to 
\begin{equation}
\begin{split}
\hspace{0.5cm}&\mathcal{M}_{\text{BLP}} \\
&= \underset{n_{1},n_{2},n_{3}} {\mathrm{max}}\frac{1}{2}(C^{(6)}_{X=3}(\infty) - C^{(6)}_{X=3}(T_{-}^{(6), X=3})) \sqrt{n_{1}^{2}+n_{2}^{2}+n_{3}^{2}}
\end{split}
\end{equation}
where $n_{1}, n_{2}, n_{3}$ are the Bloch vectors of the qubit state $\rho$. Since $\sqrt{n_{1}^{2}+n_{2}^{2}+n_{3}^{2}}$ can have a maximum value of $1$, so $\mathcal{M}_{\text{BLP}}$ for this \texttt{SWITCH} configuration is $$\frac{1}{2} [C^{(6)}_{X=3}(\eta=0)-C^{(6)}_{X=3}(\eta=\frac{2}{3})] = \frac{1}{3}=0.333 $$
Therefore, the normalized BLP measure becomes $0.250$.

Similarly, $\mathcal{M}_{\text{BLP}}$ for $3$-\texttt{SWITCH} with $3$ channels ($n=3$, $X=3$) turns out to be $\frac{1}{2} [C^{(3)}(\eta=0)-C^{(3)}(\eta=0.342)] = 0.070$ and the normalized measure is $0.065$.
This implies, that the normalized BLP measure is higher for the full-\texttt{SWITCH} (having $6$ dimensional control) with $3$ channels than for $3$-\texttt{SWITCH} with $3$ channels. This again justifies that the dimension of control behaves as an important resource behind the enhancement in \texttt{SWITCH} performance.
 
\end{document}